\documentclass{elsarticle}

\usepackage{hyperref}
\usepackage{graphicx}
\usepackage{color}
\usepackage{url}
\usepackage{mathtools}
\usepackage{multirow}
\usepackage{listings}
\usepackage{paralist}
\usepackage{listings}
\usepackage{enumitem}
\usepackage{array}
\usepackage{supertabular}

\journal{Journal of \LaTeX\ Templates}

\newcommand\approach{\emph{PreMiSE}~}
\newcommand\approachNoSpace{\emph{PreMiSE}}
\definecolor{review}{rgb}{1.0, 0.11, 0.0}

\newenvironment{change}{\color{black}}{\color{black}}

\newenvironment{review}{\color{black}}{\color{black}}

\begin{document}

\begin{frontmatter}

\title{Predicting Failures in Multi-Tier Distributed Systems}

\author[1]{Leonardo Mariani}
\author[1,2]{Mauro Pezz\`e}
\author[1]{Oliviero Riganelli\corref{oliviero}}
\ead{oliviero.riganelli@unimib.it}
\author[2]{Rui Xin}

\address[1]{Universit\`a di Milano-Bicocca, Milan, Italy}
\address[2]{Universit\`a della Svizzera italiana (University of Lugano), Lugano, Switzerland}

\cortext[oliviero]{Corresponding author}




\begin{abstract}
Many applications are implemented as multi-tier software systems, and are executed on distributed infrastructures, like cloud infrastructures, to benefit from the cost reduction that derives from dynamically allocating resources on-demand.
In these systems, failures are becoming the norm rather than the exception, and predicting their occurrence, as well as locating the responsible faults, are essential enablers of preventive and corrective actions that can mitigate the impact of failures, and significantly improve the dependability of the systems. 
%
%
\begin{change}Current failure prediction approaches suffer either from false positives or limited accuracy, and do not produce enough information to effectively locate the responsible faults.\end{change}

In this paper, we present \approachNoSpace, a lightweight and precise approach to predict failures and locate the corresponding faults in multi-tier distributed systems. 
\approach blends anomaly-based and signature-based techniques to identify multi-tier failures that impact on \begin{change}performance indicators, with high precision and low false positive rate\end{change}.
%
The experimental results that we obtained on a Cloud-based IP Multimedia Subsystem indicate that \approach can indeed predict and locate possible failure occurrences with high precision and low overhead.


\end{abstract}

\begin{keyword}
Failure prediction\sep Multi-tier distributed systems\sep Self-healing systems\sep Data analytics\sep Machine learning\sep Cloud computing

\end{keyword}

\end{frontmatter}


\section{Introduction}
\label{sec:introduction}
\begin{change}Multi-tier distributed systems
are systems composed of several distributed nodes organized in layered tiers. 
Each tier implements a set of conceptually homogeneous functionalities that provides services to the tier above in the layered structure, while using services from the tier below in the layered structure.  
The distributed computing infrastructure and the connection among the vertical and horizontal structures make multi-tier distributed systems extremely complex and difficult to understand even for those who developed them.

Indeed, runtime failures are becoming the norm rather than the exception in many multi-tier distributed systems, such as ultra large systems~\cite{Feiler:ultraLargSystems:cmu:2006} systems of systems~\cite{Sommerville:SoS:cacm:2012,Nielsen:SoS:ACM-CR:2015} and cloud systems~\cite{Chen:cloudFailures:ISSRE:2014, Ko:cloudFailures:IEEESpec:2012, Vishwanath:cloudFailures:SoCC:2012}. In these systems, failures become unavoidable due to 
both their characteristics and the adoption of commodity hardware.\end{change}
The characteristics that increase the chances of failures are the increasing size of the systems, the growing complexity of the system--environment interactions, the heterogeneity of the requirements and the evolution of the operative environment. 
The adoption of low quality commodity hardware is becoming common practice in many contexts, notably in cloud systems~\cite{ETSI:NFV:2014, Bauer:CloudReliability:2012}, and further reduces the overall system reliability.

Limiting the occurrences of runtime failures is extremely important in many common applications, where runtime failures and the consequent reduced dependability negatively impact on the expectations and the fidelity of the customers, and becomes a necessity in systems with strong dependability requirements, such as telecommunication systems that telecom companies are migrating to cloud-based solutions~\cite{ETSI:NFV:2014}.

\emph{Predicting failures at runtime} is essential to trigger automatic and operator-driven reactions to either avoid the incoming failures or mitigate their effects with a positive impact on the overall system reliability.
Approaches for predicting failures have been studied in several contexts, such as mobile devices~\cite{Riganelli:PowerOptimization:HASE:2008,Nistor:SunCat:ISSTA:2014}, system performance deviations~\cite{Malik:AutomaticDetection:ICSE:2013,Jin:Performance:2007}, distributed systems~\cite{Williams:BlackBoxPrediction:IPDPS:2007,Fu:EventCorrelation:SC:2007}, online and telecommunication applications~\cite{Ozcelik:Seer:TSE:2016,Salfner:PredictingFailure:IPDPS:2006,Haran:ApplyingClassification:FSE:2005}. 

Current approaches for predicting failures exploit either anomaly- or signature-based strategies. 
%
%
Anomaly-based strategies consider behaviors that significantly deviate from the normal system behavior as symptoms of failures that may occur in the near future~\cite{Sauvanaud:Anomaly:ISSRE:2016,Jin:Performance:2007,Guan:failurePrediction:ICCCN:2011,Williams:BlackBoxPrediction:IPDPS:2007,Tan:AnomalyPrediction:PODC:2010,Tan:anomalyPrediction:ICDCS:2012,IBM:SmartCloudAnalyticsPI:2015}. Anomaly-based techniques suffer from false positives, because of the difficulty of distinguishing faulty from rare albeit legal behaviors, in the absence of information about failure patterns.

\begin{change}Signature-based strategies rely on known patterns of failure-prone behaviors, called signatures, to predict failures that match the patterns~\cite{Ozcelik:Seer:TSE:2016,Nistor:SunCat:ISSTA:2014,Malik:AutomaticDetection:ICSE:2013,Vilalta:EventPrediction:ICDM:2002,Fu:EventCorrelation:SC:2007,Salfner:PredictingFailure:IPDPS:2006}. 
By working with known patterns, signature-based techniques cannot cope with emerging failures. 
Moreover, signature-based techniques usually
work with patterns of discrete events, 
such as error reports and system reboots, and do not cope well with failures that directly impact on performance indicators whose values vary in continuous domains over time.
Performance indicators with continuous variables that span a wide range of values are common in multi-tier distributed systems, and signature-based techniques working on simple sample-based discretization often have limited accuracy in the presence of combinations of values not experienced in the past.\end{change}

\begin{figure}[th!]
\begin{center}
  \includegraphics[width=1.0\columnwidth]{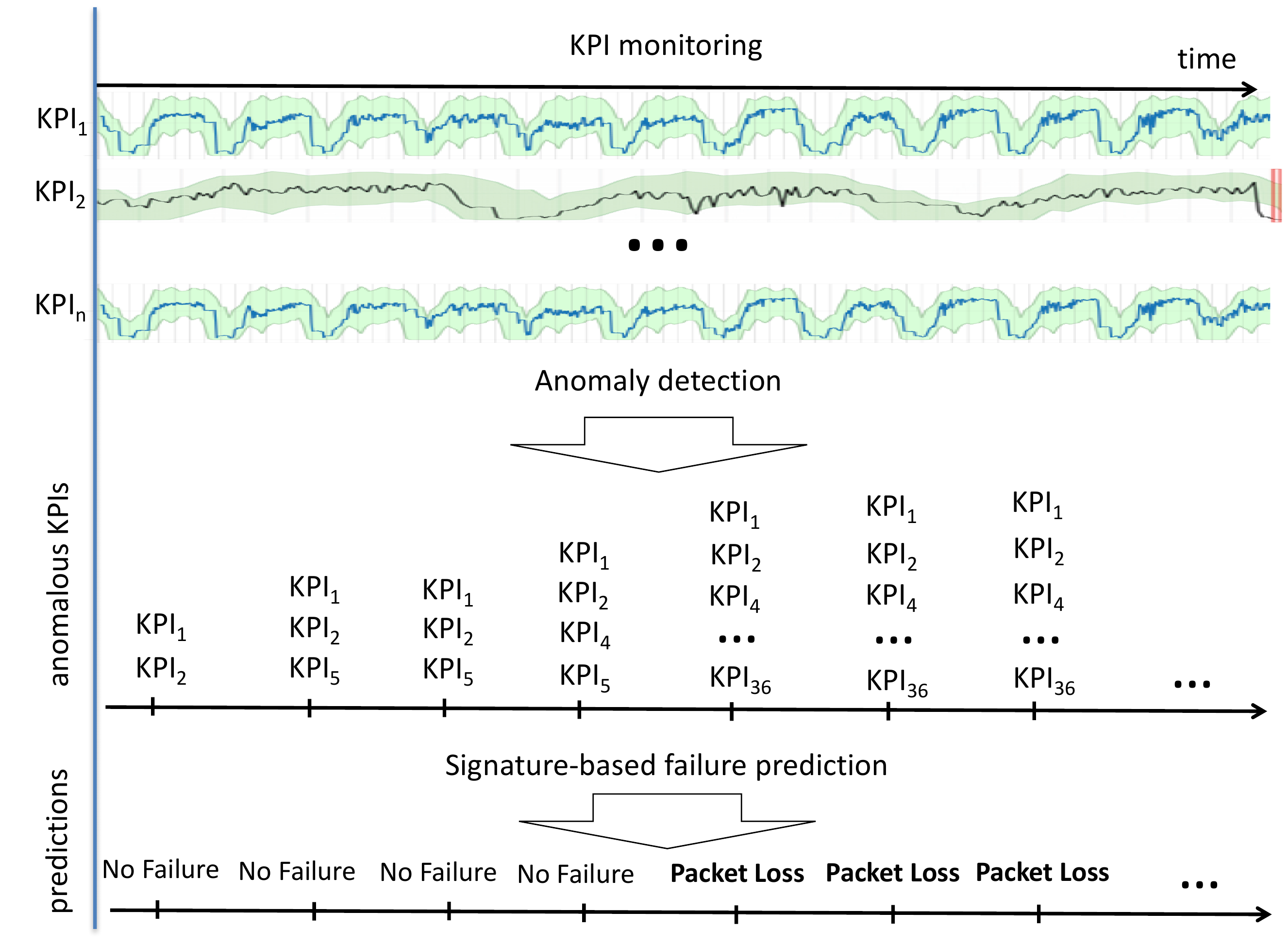}
\caption{The overall flow of \approach online activities to predict failures}
\label{fig:intuitionProud}
\end{center}
\end{figure}

In this paper, we present \approach (PREdicting failures in Multi-tIer distributed SystEms), a novel approach that can accurately predict failures and precisely locate the responsible faults in \begin{change} multi-tier distributed systems.
By addressing the challenges that characterize complex multi-tier distributed systems, \approach addresses also the subset of challenges that characterize singe-tier systems. 
\end{change} 
\approach originally combines signature-based with anomaly-based approaches, to improve the accuracy of signature-based approaches in predicting failures that impact on \begin{change}  performance indicators\end{change}. 
As illustrated in Figure~\ref{fig:intuitionProud}, \approach
\vspace{-0.3cm}
\begin{enumerate}[label=-,itemsep=3pt]

\item monitors the status of the system \begin{change}by collecting\end{change} (a large set of) performance indicators from the system nodes, for instance \emph{CPU utilization} for each CPU in the system, that we refer to as Key \begin{change}Performance\end{change} Indicators (KPIs) (\emph{KPI monitoring} in the figure),  
\begin{change} 
\item  identifies deviations from normal behaviors by pinpointing anomalous KPIs with anomaly-based techniques 
(\emph{Anomaly detection} in the figure),

\item identifies incoming failures by identifying symptomatic anomalous KPI sets with signature-based techniques.\end{change}
(\emph{Signature-based failure prediction} in the figure).

\end{enumerate}

\medskip

In the \emph{KPI monitoring} activity, \approach collects KPIs from different layers of the target multi-tier distributed system. \begin{change} KPIs are metrics collected on specific resources, and are the performance indicators that failure prediction approaches use to  estimate the status of the system.\end{change}
In the \emph{anomaly detection} activity, \approach exploits multivariate time series analysis to identify anomalies.
In details, \approach elaborates the KPI values collected during a training phase to produce a baseline model that represents the legal behavior of the system, and relies on time series analysis to combine the information from multiple KPIs provided in the baseline model for revealing anomalous behaviors. 
For example the baseline model can identify a simultaneous increase in both memory usage and memory cached as either a symptom of an anomalous behavior when occurring in the presence of a normal workload, or as a normal albeit infrequent behavior when occurring in the presence of a high workload. 
The baseline model accurately reveals anomalies in the behavior of the system as a whole, but cannot 
\begin{inparaenum}[(i)] 
\item distinguish between malign and benign anomalies, that is, symptoms of incoming failures from normal albeit uncommon behaviors, 
\item predict the type of the incoming failures, and 
\item locate the sources of the incoming failures.
\end{inparaenum}

In the \emph{failure prediction} activity, \approach exploits signature-based techniques to accurately distinguish malign from benign anomalies:
It identifies the incoming failures that correspond to malign anomalies, predicts the type of  incoming failures, and locates the sources of incoming failures.
More in details, \approach uses historical data about correct and failing behaviors to learn patterns that correlate malign anomalies to failure types, and to relate failures to failure sources.
For example, the signature-based failure prediction activity may discard as benign series of anomalous combination of memory usage, memory cached and normal workload, and identify an excessive re-transmission of network packets jointly with a lack of system service response as symptoms of a possible packet loss problem in a network node, problem that may cause a system failure in the long run.

We evaluated \approach 
on a prototype multi-tier distributed architecture that implements telecommunication services.  
The experimental data indicate that \approach can predict failures and locate faults with high precision and low false positive rates for some relevant classes of faults, thus confirming our research hypotheses.

The main contributions of this paper are: 

\begin{enumerate}[label=-,itemsep=3pt]

\item An approach that combines anomaly- and signature-based techniques to predict failure occurrences and locate the corresponding faults with high precision and low false positive rates, by exploiting information collected from performance indicators in multi-tier distributed systems. 
The proposed \approach approach can distinguish between anomalous albeit legal behaviors from erroneous behaviors that can lead to failures, and can identify the type and location of the causing faults.

\item A set of experimental results obtained on a multi-tier distributed system that hosts a telecommunication system, which resembles an industrial telecommunication infrastructure, and which provides evidence of the  precision and accuracy of the approach in the context of cloud systems, a relevant type of multi-tier distributed systems.

\end{enumerate}

The paper is organized as follows. 
Section~\ref{sec:systemDesign} introduces the \approach approach. 
Section~\ref{sec:offline} discusses the offline training of the models.
Section~\ref{sec:online} presents the online failure prediction mechanism, based on an original combination of anomaly- and signature-based techniques.
Section~\ref{sec:evaluationMethodology} illustrates the methodology that we followed to evaluate the approach, introduces the evaluation metrics and the experimental setting, provides the essential implementation details of the evaluation infrastructure, and presents both the types of faults injected in the system and the reference workload used to evaluate the approach. 
 Section~\ref{sec:results} discusses the experimental results about the effectiveness and the overhead of the proposed approach. 
Section~\ref{sec:related} overviews the main related approaches, highlighting the original contribution of our approach.
Section~\ref{sec:conclusion} summarizes the main contribution of the paper, and indicates the research directions open with the results documented in this paper. 

\section{The \approach approach}
\label{sec:systemDesign}

\approach detects failure symptoms, correlates the detected symptoms to failure types, and locates the resources responsible of the possible failures which may occur in the near future. 

Several anomalous behaviors of many types can often be observed well in advance with respect to system failures, which can be frequently mitigated or avoided, especially in multi-tier distributed systems. 
For instance in cloud systems, early observed communication issues can trigger dynamic reallocation of resources to mitigate or avoid failures. 
%
Differently from current approaches, which simply report anomalous behaviors~\cite{Tan:anomalyPrediction:ICDCS:2012,Dean:anomalyPrediction:ICAC:2012,Tan:AnomalyPrediction:PODC:2010,Guan:failurePrediction:ICCCN:2011}, \approach 

\begin{enumerate}[label=-,itemsep=3pt]

\item distinguishes anomalous behaviors that are caused by software faults and that can lead to failures from anomalous behaviors that are derived from exceptional albeit legal situations and that do not lead to failures, thus reducing the amount of false alarms of current failure prediction approaches,

\item correlates anomalous behaviors detected at the system level to specific types of faults, and predicts not only the occurrence but also the type of possible failures,
thus simplifying the identification of effective corrective actions, and

\item identifies the resources likely responsible for the predicted failure, thus providing the developers with a useful starting point for investigating and solving the problem.

\end{enumerate}

\medskip

%

\begin{figure*}[ht!]
 \begin{center}
   \includegraphics[width=1.0\columnwidth]{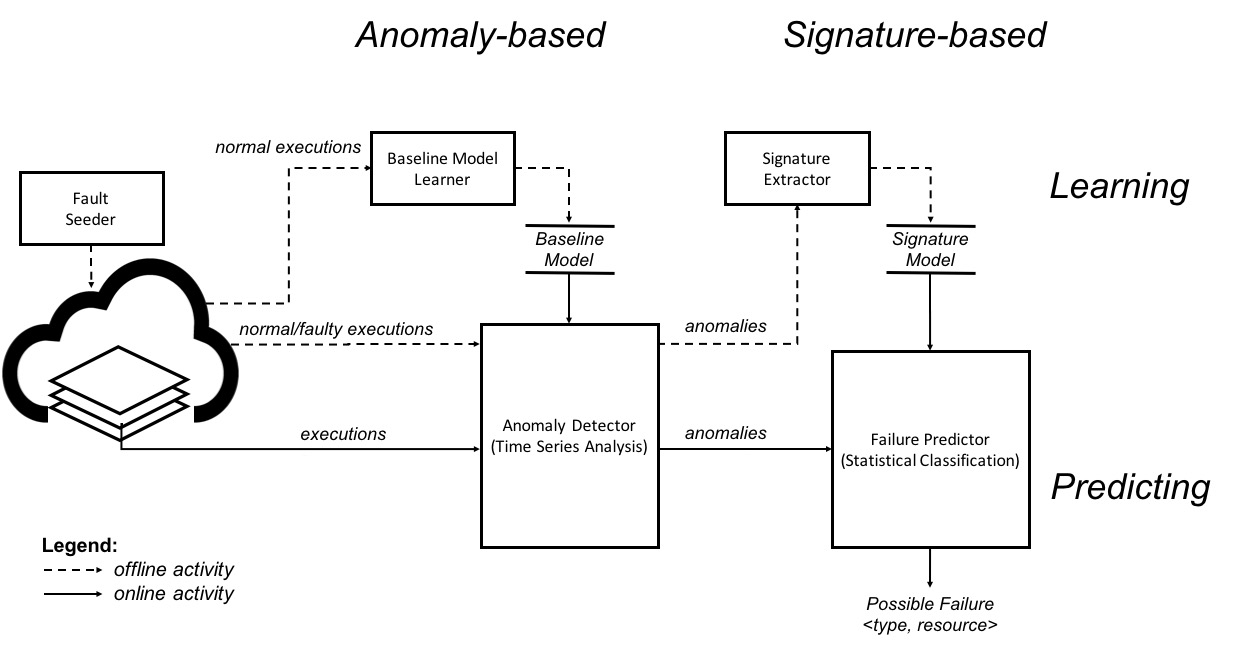}
   \caption{The \approach learning and predicting phases}
 \label{fig:onlinePredicting}
 \end{center}
 \end{figure*}

As illustrated in Figure~\ref{fig:onlinePredicting}, \approach is composed of an \mbox{offline} model training and an online failure prediction phase. 
As discussed in details in the next sections, in the \emph{offline model training} phase, \approach builds baseline and signature models that capture the system behavior, and in the \emph{online failure prediction} phase, \approach uses the baseline and signature models to detect anomalies and predict failures, respectively. 

%
%

\section{Offline Model Training}
\label{sec:offline}


In the offline learning phase \approach builds a \emph{baseline model} and a \emph{signature model}. The baseline model identifies anomalous behaviors that might be interpreted as symptoms of failures, while the \emph{signature model} associates sets of anomalous behaviors to either legal albeit spurious behaviors or symptoms of future failures, and locate the resources likely responsible for the failure.

As illustrated in Figure~\ref{fig:onlinePredicting} in the offline learning phase, \approach \emph{monitors series of KPIs} over time under normal execution conditions to \emph{learn the baseline model}, and \emph{seeds faults} of the target types to \emph{extract the signature model}.  

\begin{review} The baseline model is a system model and, as such, it is obtained by modeling only the failure-free behavior, that is, the normal behavior of the system. The model is used to calculate the expected values, at which the measured current values are compared to. If the expected and actual values differ significantly, the system is suspected to not behave as intended. The detection of several anomalous values is a relevant indicator of failures that may happen in the future.

In contrast with the baseline model, which focuses on failure-free behaviors, the generation of the \emph{signature model} requires training data for both the failure-prone and failure-free executions. \approach uses the \emph{signature model} to decide whether sets of anomalies are due to novel but normal behaviors or specific classes of failures. \end{review} 


%
\approach can build the models from different kinds of KPIs, granted that their values can be monitored as series of values over time, and can train the signature model with different kinds of seeded faults, granted that the consequent failures occur after some system degradation over time.

As a simple example, \approach might detect combinations of anomalous rates of packet re-transmission and aborted operations, by means of the baseline model.
It may then associate these anomalous behaviors to either a transient situation due to a high and unexpected peak of requests or to a communication problem that will likely cause a system failure in the future, by means of the signature model.
It may also identify the subsystems likely responsible for the incoming communication problems, from the information provided with the detected violations patterns.

\begin{review}While model incompleteness is both possible and probable, this can be compensated by incrementally collecting additional evidence about the behavior of the system. For instance, the baseline model can be improved incrementally and the signature model can be retrained regularly.\end{review}

\subsection{KPI Monitor}

\approach collects an extensive number of KPIs from different tiers of many system components without interfering with the system behavior by relying on lightweight monitoring infrastructures, often natively available at the different levels~\cite{Case:SNMP:RFC:1996, Openstack:Ceilometer:2015}, and elaborates the collected data on an independent node that executes the data processing routines. 
In this way \approach affects the monitored nodes only with the negligible costs of collecting data, and not with the relevant costs of the computation, which is relegated to an independent node. 
%
%
The empirical results reported in Section~\ref{subsec:RQ5} confirm the non-intrusiveness of the monitoring infrastructure. 


The values monitored for each KPI are time series data, that is, sequences of numeric values, each associated with a timestamp and a resource of the monitored system. Table~\ref{tab:timeSeriesBytesSentPerSec} reports a sample excerpt of the time series data for the KPI \emph{BytesSentPerSec} collected from a resource named \emph{Homer}\footnote{Resource \emph{Homer} is one of the virtual machines used in our empirical setting, a protoype of a  cloud infrastructure used by telecommunication companies to provide SIP services, as described in Section~\ref{sec:implementationDetails}. 
Resource \emph{Homer} is a standard XML Document Management Server that stores MMTEL (MultiMedia TELephony) service settings of the users.}. 
Columns \emph{Timestamp}, \emph{Resource} and \emph{BytesSentPerSec} indicate the time information, the name of the resource that produced the KPI value, and the number of bytes sent in the last second, respectively. 

\begin{review}In the context of this paper, \end{review}KPIs are metrics measured on specific resources.  More formally, a KPI is a pair $<resource, metric>$.  
For example, \emph{BytesSentPerSec} collected at \emph{Homer} is a KPI, and the same metric collected at another resource is a different KPI.  
Thus, the number of collected KPIs depends on the product of monitored metrics and resources.

\approach can be customized with different sets of KPIs, collected at heterogeneous levels. The selection of KPIs depends on the characteristics of the target system, and impacts on the type of failures that \approach can detect:  \approach can detect only failures whose occurrence follows some anomalous behaviors that reflect in  anomalies in the series of KPIs monitored over time. 
We experimented with over 600 KPIs of over 90 different types, collected KPIs at different abstraction levels, ranging from the Linux operating system to the Clearwater application level, and predicted failures of different types, ranging from network to memory failures. 
We discuss the experimental setting and the results in details in Section~\ref{sec:implementationDetails}.


\begin{table}[h]
\caption {Sample time series for KPI \emph{BytesSentPerSec} collected at node \emph{Homer}}
\label{tab:timeSeriesBytesSentPerSec}
\begin{center}
\begin{tabular}{|c|c|c|}
\hline
\textbf{Timestamp} & \textbf{Resource} & \textbf{BytesSentPerSec}\\
\hline

\ldots & \ldots & \ldots \\

Dec. 20, 2016 22:22:35 &
Homer &
101376 \\

Dec. 20, 2016 22:23:36 &
Homer &
121580 \\

Dec. 20, 2016 22:24:36 &
Homer &
124662 \\

Dec. 20, 2016 22:25:36 &
Homer &
106854 \\

\ldots & \ldots & \ldots \\

\hline
\end{tabular}
\end{center}
\end{table}

\subsection{Baseline Model Learner}

The \emph{baseline model learner} derives the baseline model from series of KPIs collected under normal system executions, and thus represents correct system behaviors.  
%
The \emph{baseline model learner} generates models with inference solutions that capture temporal relationships in time series data, by including trends and seasonalities~\cite{Box:TimeSeriesAnalysis:2008}. 
In particular, the \emph{baseline model learner} applies Granger causality tests~\cite{Arnold:TCMGranger:KDD:2007} to determine wether a time series variable can predict the evolution of another variable. 


A time series $x$ is said to be a \emph{Granger cause} of a time series $y$, if and only if the regression analysis of $y$ based on past values of both $y$ and $x$ is statistically more accurate than the regression analysis of $y$ based on past values of $y$ only. 
%
%
The ability of Granger causality analysis to analyze the dependency between KPIs is a key factor for improving the accuracy of the analysis, because many KPIs are correlated. 
For instance the CPU load often depends on the rate of incoming requests, and several phenomena could be fully interpreted only by considering multiple time series jointly. 
For instance, a high CPU load might be anomalous or not depending on the rate of incoming requests that are received by the system. 

%

\begin{figure}
\begin{center}
\includegraphics[width=12cm]{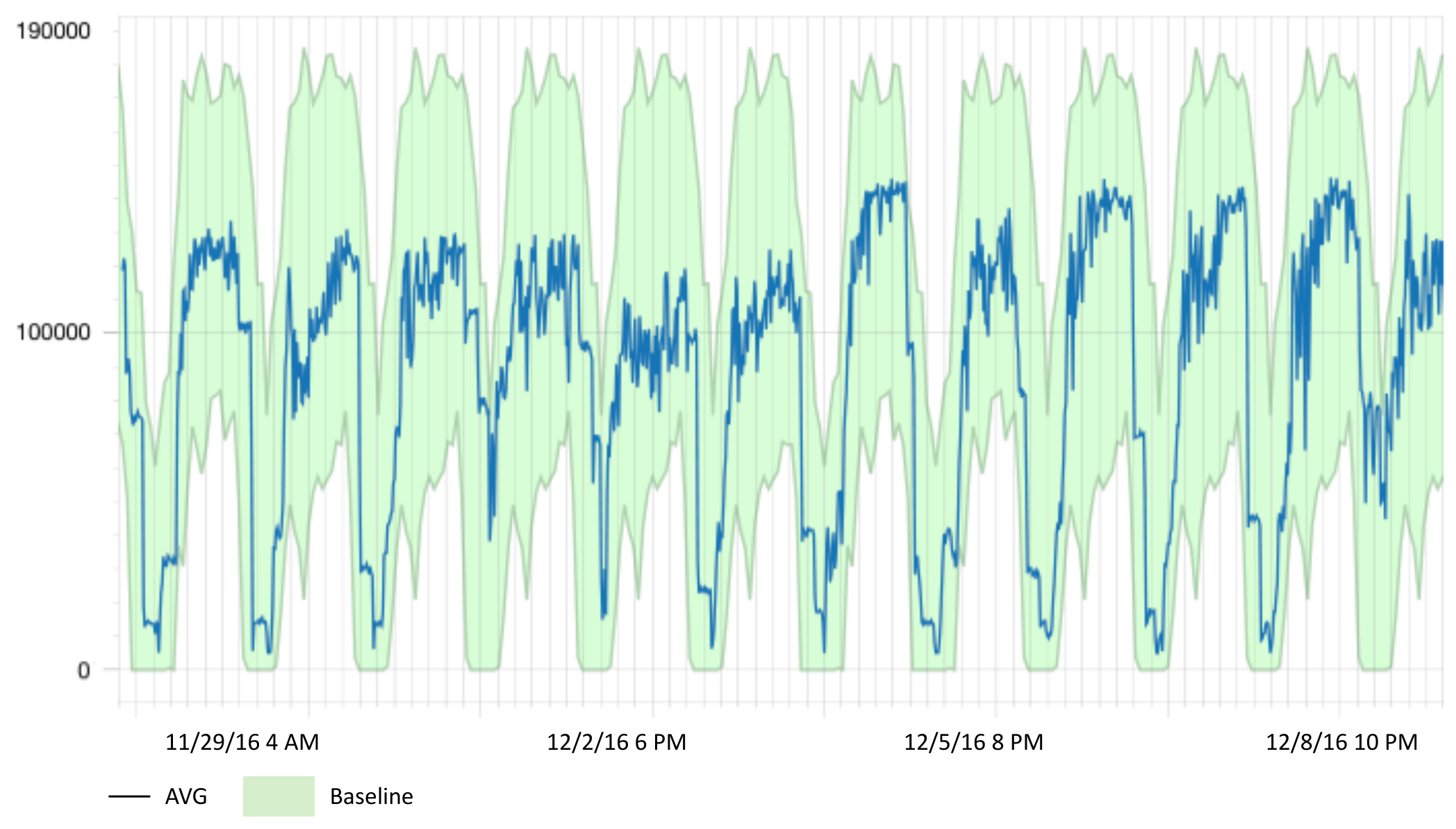}

\caption{A sample baseline model of a single KPI: 
The \emph{BytesSentPerSec} KPI for the \emph{Homer} virtual machine}
\label{fig:baselineModel}
\end{center}
\end{figure}
\begin{figure}
\begin{center}
\includegraphics[width=7cm]{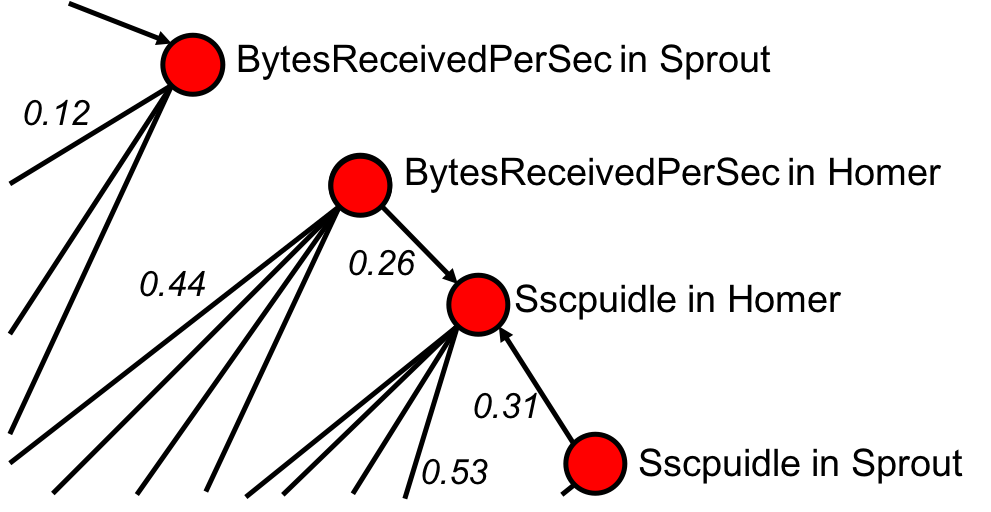}

\caption{A baseline model of the correlation among KPIs: an excerpt from a Granger causality graph}
\label{fig:casualityGraph}
\end{center}
\end{figure}

The \approach baseline model includes both models for the single KPIs and models of the correlation among KPIs. 
Figure~\ref{fig:baselineModel} illustrates a baseline model of a single KPI, namely the model inferred for KPI \emph{BytesSentPerSec} collected from the \emph{Homer} virtual machine. The figure indicates the average value of the time series (dark blue line) and the confidence interval for new samples (light green area around the line). 
Figure~\ref{fig:casualityGraph} shows an excerpt of a Granger causality graph that represents the causal dependencies among KPIs. Causal dependencies indicate the strength of the correlation between pairs of KPIs, that is, the extent to which changes of one KPI are related to changes of another KPI. Nodes in the causality graph correspond to KPIs, and weighted edges indicate the causal relationships among KPIs, as a value in the interval $[0,1]$, indicating the increasing strength of the correlation. In the example, the values of \emph{BytesSentPerSec} metric in node \emph{Homer} are strongly correlated to and can thus be used to predict values of \emph{Sscpuidle} metric in node \emph{Homer}. 


\subsection{Fault Seeder}

The baseline model captures both the shape of the KPI values over time for single KPIs and the correlation among KPIs under normal execution conditions. 
The signature model captures the relations among anomalous KPIs observed both during normal execution conditions and with seeded faults. 
The Signature Model Extractor can be trained with different types of seeded faults, whose consequent failures occur after some system degradation over time. Being trained with seeded faults, the signature model can build patterns of anomalies related to failures, and thus distinguish between benign and failure-prone anomalies when monitoring the system execution.
%
%
%
The  \emph{Fault seeder} injects a fault of a given type in a system location for each system run. 
%
%
\approach can accurately predict failures and locate faults of the types of faults injected in the training phase.  Following  common practice, we chose the faults to inject according to the Pareto distribution of the frequency and severity of the fault types.
%
%
The signature model can be extended incrementally to new types of faults.

\subsection{Signature Model Extractor} 
\label{subsec:learningFPModel}


The \emph{Signature model extractor} derives the signature model from a sample set of anomalous behaviors that correspond to correct as well as faulty executions. 
The \emph{Signature model extractor} learns the model from anomalies identified by the Anomaly Detector\footnote{We discuss the Anomaly Detector in details in Section~\ref{sec:online} reveals during the training phase, by relying on the baseline model and faulty execution traces.}. Anomalies are tuples $\langle (a_1,\ldots, a_n), f, r \rangle$, where $(a_1,\ldots, a_n)$ is a (possibly empty) sequence of anomalous KPIs that are detected during an execution window of a fixed length, $f$ is a failure type, and $r$ is the resource responsible for the failure. Thus, anomalous KPIs are KPIs without time stamp, and indicate that the KPIs assume anomalous values for at least a time stamp within the considered window.
For instance, the tuple $\langle$($\langle BytesReceivedPerSec, Homer\rangle$, $\langle Sprouthomerlatencyhwm, Sprout\rangle$, 
$\langle Sscpuidle, Sprout\rangle$), $Packet loss, Sprout\rangle$ indicates three correlated KPIs that assume anomalous values in the considered execution window, and signals a predicted packet loss failure in node Sprout with a likelihood encoded in the tuple $\langle 30,1 \rangle$, as discussed after in this section.

Both $f$ and $r$ are empty when the execution window corresponds to a normal execution with no active faults.
In the training phase \approach seeds at most a fault per execution window, and considers execution windows that originate with the activation of the fault and slide through the faulty execution up to a maximum size, and thus collects anomalies that occur immediately after the fault activation as well as along a possibly long lasting failing execution.  
We discuss the tuning of the size of the sliding window in Section~\ref{sec:results}. 

\approach records both the type of the failure and the resource seeded with the fault to learn signatures that can predict the failure type and locate the responsible resource.
The \emph{Signature Model Extractor} relies on several faults for each type, seeded in different locations, and uses multi-label probabilistic classifiers as signature extractors. 

Probabilistic classifiers generate probability distributions over sets of class labels from given sets of samples. 
\approach uses the probability distributions to compute the confidence of the predictions, thus producing a signature model that the \emph{Failure predictor} can exploit to predict both the type of the failure and the location of the resources that are most likely responsible for the failure, and to compute the confidence on the prediction.
We empirically investigated signature extractors based on \emph{Support Vector Machine (SVM)}, \emph{Bayesian Network (BN)}, \emph{Best-First Decision Tree (BFDT)}, \emph{Na\"ive Bayes (NB)}, \emph{Decision Table (DT))}, \emph{Logistic Model Tree (LMT)} and \emph{Hidden Na\"ive Bayes (HNB)} algorithms.
We introduce the algorithms in Section~\ref{sec:implementationDetails}, and discuss their experimental evaluation in Section~\ref{sec:results}.
%

\begin{figure}
 \begin{center}
  \includegraphics[width=9.5cm]{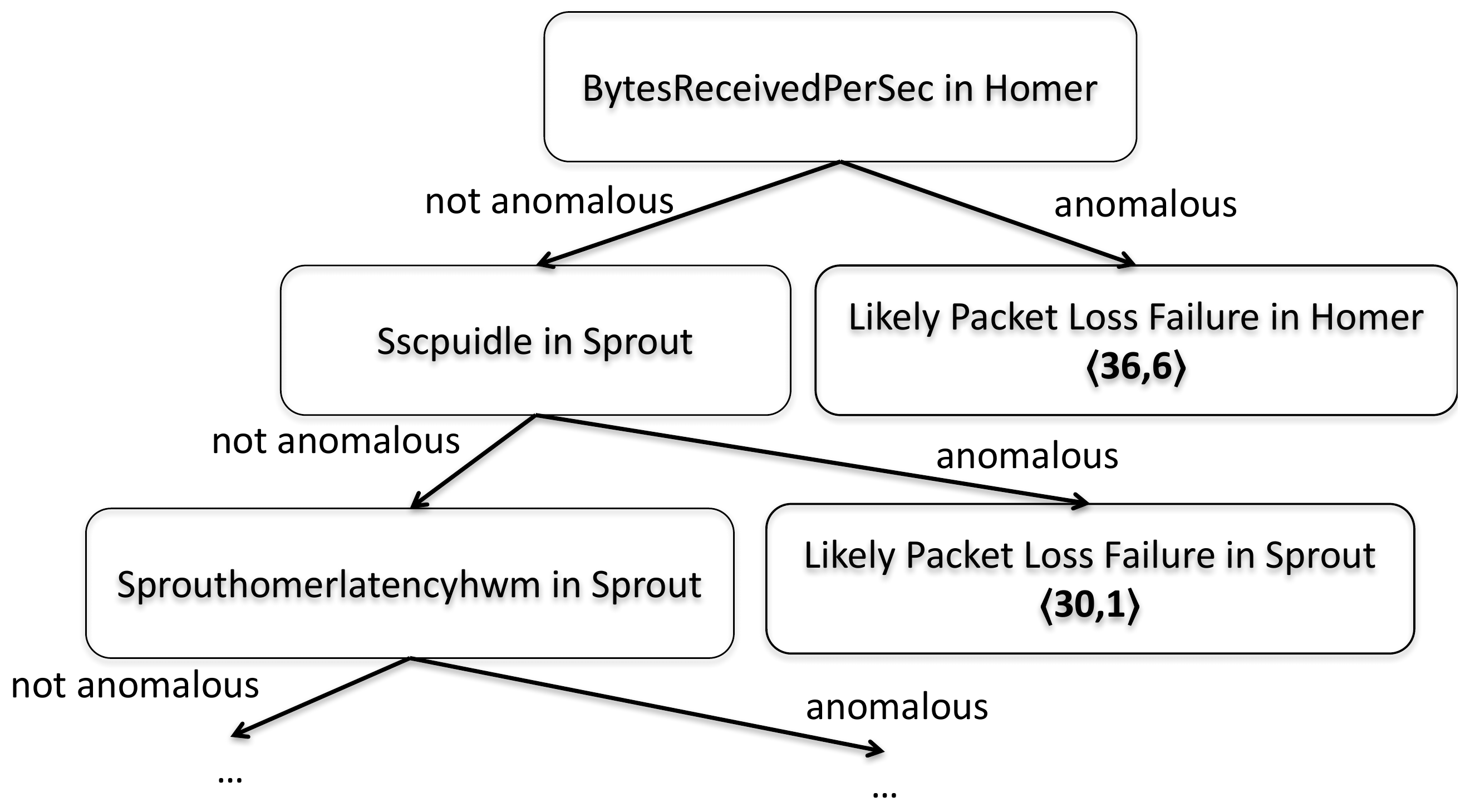}
 \caption{A sample signature model based on decision trees}
 \label{fig:decisionTree}
 \end{center}
 \end{figure}

As an example of signature model, Figure~\ref{fig:decisionTree} shows an excerpt of a decision tree that \approach inferred for packet loss failures\footnote{We experimented with different models.  Here we report a signature in the form of a decision tree because decision trees are easier to visualise and discuss than other models. 
}.
%
%
Nodes correspond to KPIs and edges indicate the \emph{anomaly} relation. Leaf nodes are annotated with the confidence level of the prediction indicates as pairs $\langle total, correct\rangle$, where \emph{total} is the amount of samples that reach the node, and \emph{correct} is the amount of samples that are correctly classified according to the leaf, that is, the number of samples corresponding to failed executions of the specific type caused by the specific resource, as indicated in the model.  The ratio between \emph{correct} and \emph{total} indicates the likelihood of the prediction to be correct.  

The model indicates that  anomalous values of \emph{BytesReceivedPerSec} in node \emph{Homer} are a symptom of a possible packet loss failure in node \emph{Homer}, that a combination of non-anomalous values of \emph{BytesReceivedPerSec} in node \emph{Homer} with anomalous values of \emph{Sscpuidle} in node \emph{Sprout} are a symptom of a possible packet loss failure in node \emph{Sprout}, and that the likelihood of a failure increases when both \emph{BytesReceivedPerSec} in \emph{Homer} and \emph{Sscpuidle} in  \emph{Sprout} are anomalous.
This may happen because packet loss problems may cause a drop in the number of accesses to the user service settings stored in the \emph{Homer} XDMS server,  since a packet loss problem may decrease the frequency of authentication requests received by \emph{Sprout} and thus increasing the CPU idle time. 
The branches of the decision tree not reported in the figure indicate additional relationship between symptoms and failures.


\section{Online Failure Prediction}
\label{sec:online}


In the online failure prediction phase, \approach uses the baseline model to detect anomalies and the signature model to predict failures.

\subsection{Anomaly Detector}
\label{sec:anomalyDetector}

Anomalies are behaviors that differ from the expectations, and are thus suspicious. The baseline model encodes expected behavior as a collection of time series of single KPIs and as the Granger correlation among KPIs, as illustrated in Figures~\ref{fig:baselineModel} and~\ref{fig:casualityGraph}, respectively.
The \emph{Anomaly Detector} signals univariate and multivariate anomalies when the values of the collected or correlated KPIs differ enough from the  baseline model.
Univariate anomalies depend on single KPIs, while multivariate anomalies depend on the combination of more than one KPI, each of which may or may not be identified as anomalous by the univariate analysis.

\begin{figure}
 \begin{center}
  \includegraphics[width=11cm]{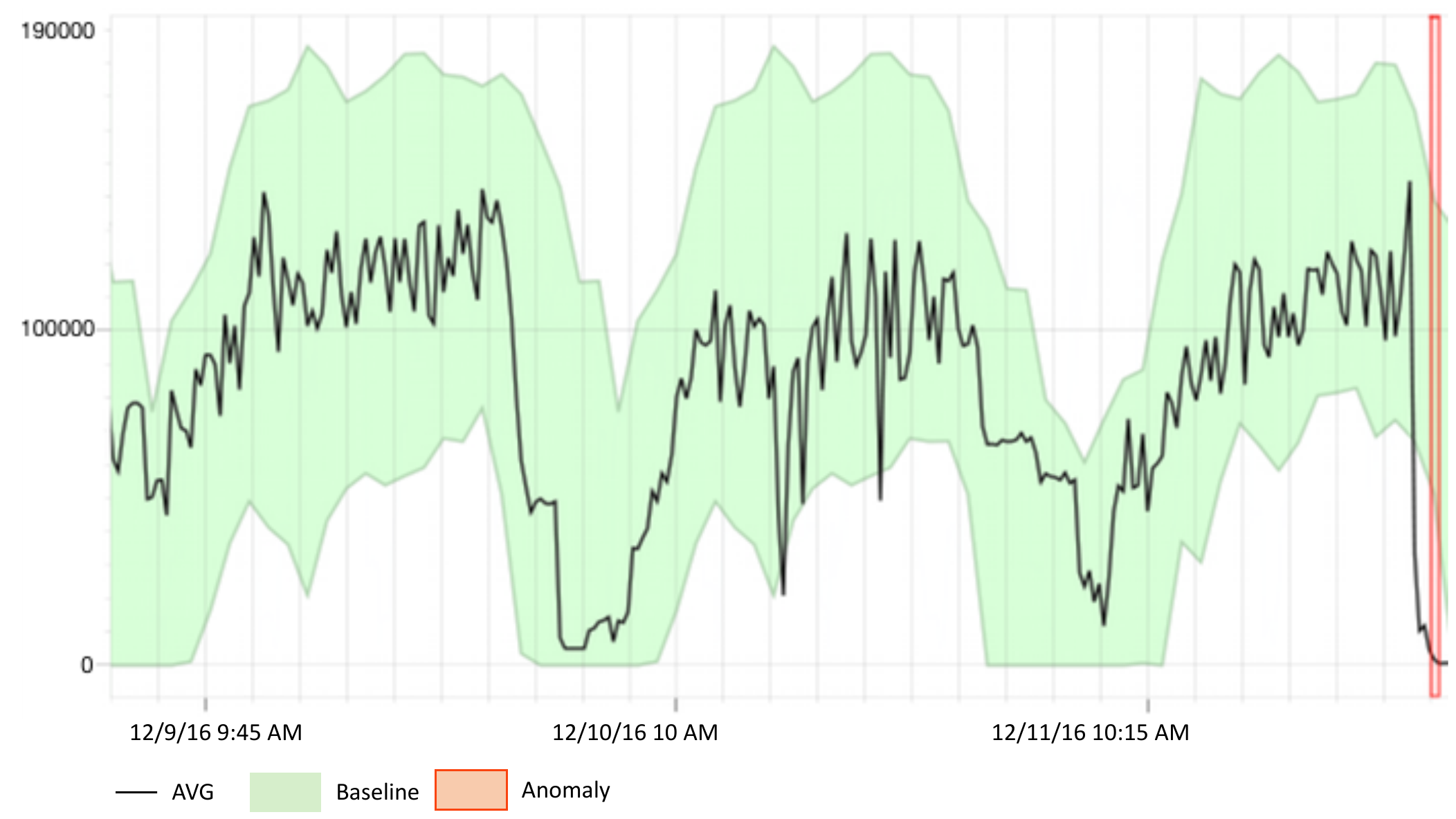}
 \caption{A sample univariate anomalous behavior}
 \label{fig:UnivariateAnomaly}
 \end{center}
 \end{figure}

The \emph{Anomaly Detector} detects univariate anomalies as samples out of range, as shown in Figure~\ref{fig:UnivariateAnomaly}. 
Given an observed value $y_t$ of a time series $y$ at time $t$, and the corresponding expected value $\hat{y}_t$ in $y$, $y_t$ is anomalous if 
the variance $\hat{\sigma}^2(y_t,\hat{y}_t)$ is above an inferred threshold.

The \emph{Anomaly Detector} detects multivariate anomalies as joint unexpected values of subsets of samples for different variables.  It deduces multivariate anomalies among the KPI variables when their relation violates the Granger correlation encoded in the Granger causality graph.   
For example, it can infer that \emph{successful call rate} and \emph{number of incoming calls} are correlated: the \emph{successful call rate} usually decreases with an increasing \emph{number of incoming calls}, and thus the anomaly detector may signal a multivariate anomaly in the presence of a decreasing \emph{successful call rate} without a corresponding increase of \emph{number of incoming calls}, regardless of the results of the univariate analysis of two values.
The \emph{Anomaly Detector} identifies multivariate anomalies with the Granger causality test that checks if a set of correlated KPIs preserves the inferred casual relationships. 


\subsection{Failure Predictor}

The \emph{Failure Predictor} identifies possible failures as sets of anomalies that find a matching in the signature model. 

As discussed in Section~\ref{subsec:learningFPModel}, \approach trains the signature model with sets of anomalies detected during execution windows of fixed size in terms of anomalous KPI samples. 
The \approach failure predictor analyzes the sets of anomalies detected in a sliding windows of the same size of the windows used during training.
%
%
For instance, the \emph{Failure Predictor} can predict an incoming packet loss failure in the presence of an anomalous value of \emph{Sscpuidle} (idle time for the authentication service) in node Sprout when occurring with a normal value of \emph{BytesReceivePerSec} (number of received requests) in the Homer XDMS server, based on the signature model shown in Figure~\ref{fig:decisionTree}.  In fact, the sequence $\langle \langle\emph{BytesReceivePerSec} \allowbreak \emph{in}~\allowbreak \emph{Homer}, \allowbreak not~anomalous\rangle, \langle\emph{Sscpuidle~in~Sprout}, anomalous \rangle\rangle$ in Figure~\ref{fig:decisionTree} leads to \emph{Likely \allowbreak Packet \allowbreak Loss \allowbreak Failure \allowbreak in \allowbreak Sprout} 

\approach generates both \emph{general} and \emph{failure-specific alerts} that correspond to generic failure-prone behaviors and specific failures, respectively.
%
%
Following common practice in failure prediction solutions that focus their capability to make prediction on recent observations~\cite{Ozcelik:Seer:TSE:2016}, \approach collects anomalies in overlapping windows sliding over time.  Anomalies first occur in the sliding window that includes the first occurrence of the anomalous KPI, and persist in the following windows, until the anomaly falls out of the windows themselves.

Right after injecting a fault, sliding windows include mostly anomalies produced during the previous failure-free execution segments and only few anomalies caused by the injected fault, while forward moving sliding windows include increasingly many  anomalies caused by the activated fault.
%
%
When sliding windows includes only a small portion of anomalies, the prediction might be imprecise. The \emph{Failure Predictor} refines the predicted failure type over time, until the prediction stabilizes. 
The \emph{Failure Predictor} copes with this transitory phase by refining an
initially general prediction into a failure-specific prediction once the
prediction stabilizes, that is, it predicts the same failure type with a
confidence level of at least 90\% for 4 consecutive times. 

In the training phase, \approach builds signature models starting with data collected just after activating the injected fault, and thus the signature model encodes early predictors, that is, sets of anomalies that occur as early as the fault is activated, often long before the consequent failure. This strategy allows \approach to quickly refine a general into a failure-specific prediction, as confirmed by the experimental results reported in Section~\ref{sec:predictionTimeResults}.

The predictions indicates the type of expected failure and the set of anomalous KPIs that substantiate the prediction.
\approach uses the information about the anomalous KPIs to localize the source of the problem, which might be a resource different from the resources responsible for the anomalous KPIs. 
For instance in our experiments, \approach correctly predicted a packet loss failure in a specific virtual machine by analyzing a set of $37$ anomalous KPIs generated by $14$ different resources. 
This is a good example of the importance of locating the fault given a large set of resources involved in the anomalous samples.

\section{Evaluation Methodology}
\label{sec:evaluationMethodology}

In this section, we introduce the research questions (Section~\ref{subsec:researchQuestions}), the testbed that implements a realistic telecom cloud-based system (Section~\ref{sec:experimentalSettings}), the prototype implementation that we used in the experiments (Section~\ref{sec:implementationDetails}), the fault seeding strategy that we adopted to collect data about failures caused by different types of faults (Section~\ref{sec:faultSeeding}), the workflow that we simulated in the experiments (Section~\ref{sec:workload}) and the quality measures that we used to evaluate the results (Section~\ref{sec:qualityMetrics}).

\subsection{Research Questions} 
\label{subsec:researchQuestions}

We address six research questions that evaluate the effectiveness of \approach, compare \approach with state of the art approaches and quantify the \approach overhead. 

\smallskip


\subsubsection*{Effectiveness}

To evaluate the capability of \approach to successfully predict failures in a realistic cloud-based system we investigate four research questions:

 \begin{description}  

\item[RQ1] \textit{Does the size of the sliding window impact on the effectiveness of \approach?} We executed \approach with different window sizes, and measured the impact of the window size on the ability to correctly predict failures. 
We used the results of this experiment to identify an appropriate window size that we used in the other experiments.  
    
\item[RQ2] \textit{Can \approach accurately predict failures and localize faults?} 
We executed \approach with different failure types and failure patterns, and measured its ability to predict failures occurrence, types and locations. 
We experimented with several multi-label mining algorithms, and compare their performance and effectiveness in predicting failures. 
We used the most effective algorithm in the other experiments. 

\item[RQ3] \textit{Can \approach correctly identify  \begin{review}normal\end{review} behaviors not experienced in the past?} 
We executed \approach with workflows that differ significantly from the workflow used in the training phase and measured its ability to classify these executions as  \begin{review}normal\end{review} executions. 

\item[RQ4] \textit{How early can \approach predict a failure?} 
We executed a number of experiments to determine how early \approach can predict failure occurrences, for different types of failures. 
\end{description}

\begin{review}RQ1 is intended to analyze the sensitivity of failure prediction and fault localization to changes in the window parameter. RQ2 focuses on the effectiveness of \approach mainly in case of faulty executions, while RQ3 studies \approach with perturbed workloads under normal conditions to study the false positive rate. RQ4 investigates prediction time in faulty executions.\end{review}

\subsubsection*{Comparison to state-of-the-art approaches}

\begin{change}[R2.1]We compare \approach with \emph{IBM OA-PI --- Operational Analytics - Predictive Insights}~\cite{IBM:SmartCloudAnalyticsPI:2015}, an industrial anomaly-based approach, and with the Grey-Box Detection Approach (G-BDA) of Sauvanaud et al.~\cite{IBM:SmartCloudAnalyticsPI:2015}, a state-of-the-art signature-based approach. We discuss the following research question:
\begin{description}
\item[RQ5] \textit{Can \approach predict failures more accurately than state-of-the-art approaches?}  
We compare \approach to OA-PI and G-BDA by executing all approaches on the same set of normal and failing executions, and comparing their ability to predict failures. 
Since OA-PI cannot predict the type of failure and locate the corresponding fault, we only evaluated its ability to predict failure occurrences.
\end{description}
\end{change}

\subsubsection*{Overhead}  

We investigated the impact of \approach on the overall performance of a cloud-based system by addressing the following research question:

    \begin{description}
    \item[RQ6] \textit{What is the overhead of \approach on the performance of the target system?} 
    This question is particularly relevant in the context of multi-tier distributed systems with strict performance requirements, like telecommunication infrastructures. 
Thus, we designed an experiment referring to such applications.  
\approach executes the resource-intensive tasks, that is, anomaly detection and failure prediction, on a dedicated physical server, and thus the overhead on the system derives only from monitoring the KPIs. 

We evaluated the impact of monitoring the KPIs on the system performance by measuring the consumption of the different resources when running the system with and without KPI monitoring active. 
Both \approach  \linebreak and data analytics solutions monitor the same KPIs, and thus share the same performance overhead, but \approach further processes the anomalies revealed with data analytics approaches, and presents more accurate predictions than competing approaches. 
\end{description}

\subsection{Testbed}\label{sec:experimentalSettings}

As representative case of multi-tier distributed system, we considered the case of a  
complete cloud-based environment running an industrial-level IP multimedia sub-system. 
To control the study, we created a private cloud consisting of \begin{inparaenum}[(i)]
\item a controller node responsible for running the management services necessary for the virtualization infrastructure, 
\item six compute nodes that run VM instances,  
\item a network node responsible for network communication among virtual machines. 
\end{inparaenum}
The characteristics of the different nodes are summarized in Table~\ref{tab:HWconf}.

\begin{table}[h]
\caption{Hardware configuration}
\label{tab:HWconf}
\centering
\begin{tabular}{|c|c|l|c|c|}
\hline
\textbf{Host} & Controller                        & Network                        & Compute (x2)                        & Compute (x4)                       \\ \hline
\textbf{CPU}  & \multicolumn{4}{c|}{\textit{\begin{tabular}[c]{@{}c@{}}Intel(R) Core(TM)2 Quad CPU Q9650\\ (12M Cache, 3.00 GHz, 1333 MHz FSB)\end{tabular}}} \\ \hline
\textbf{RAM}  & 4 GB                              & 4 GB                           & 8 GB                                & 4 GB                               \\ \hline
\textbf{Disk} & \multicolumn{4}{c|}{250 GB SATA hard disk}                                                                                                    \\ \hline
\textbf{NIC}  & \multicolumn{4}{c|}{Intel(R) 82545EM Gigabit Ethernet Controller}                                                                             \\ \hline
\end{tabular}
\end{table}

We used OpenStack~\cite{Openstack:Cloud:2015} version Icehouse on Ubuntu 14.04
LTS as open source cloud operating system
and KVM~\cite{KVM:Cloud:2015} as hypervisor. 

To evaluate our approach, we deployed Clearwater~\cite{Clearwater:CloudCaseStudy:2015} on the cloud-based infrastructure. Clearwater is an open source IP Multimedia Subsystem (IMS) and provides IP-based voice, video and message services.  Clearwater is specifically designed to massively scale on a cloud-based infrastructure, and is a product originated from the current trend of migrating traditional network functions from inflexible and expensive hardware appliances to cloud-based software solutions. Our Clearwater deployment consists of the following virtual machines:
\begin{description}
  \item [Bono:] the entry point for all clients communication in the Clearwater system.
  \item [Sprout:] the handler of client authentications and registrations.
  \item [Homestead:] a Web service interface to Sprout for
  retrieving authentication credentials and user profile information. 
  \item [Homer:]  a standard XML Document Management Server that stores
  MMTEL (MultiMedia Telephony) service settings for each user.
  \item [Ralf:]  a service for offline billing capabilities.
  \item [Ellis:] a service for self sign-up, password
  management, line management and control of multimedia telephony service
  settings.
\end{description}

Each component is running on a different VM. Each VM is configured with 2 vCPU, 2GB of RAM and 20GB hard disk space, and runs the Ubuntu 12.04.5 LTS operating system.

Our multi-tier distributed system is thus composed of eight machines running components from three tiers: the operating system, infrastructure and application tiers, running Linux, OpenStack, and virtual machines with Clearwater, respectively. We refer to this environment as the \emph{testbed}.

\subsection{Prototype Implementation}
\label{sec:implementationDetails}

Our prototype implementation of the monitoring infrastructure collects a total of 633 KPIs of 96 different types from the 14 physical and virtual machines that comprise the protoype. We collected KPIs at three levels: 162 KPIs at the application level with the SNMPv2c (Simple Network Management Protocol) monitoring service for Clearwater~\cite{Case:SNMP:RFC:1996}, 121 KPIs at the IaaS level with the OpenStack telemetry service (Ceilometer) for OpenStack~\cite{Openstack:Ceilometer:2015} and 350 KPIs at the operating system level with a Linux OS agent that we implemented for Ubuntu. 
We selected the KPIs referring to the multi-tired distributed nature of our prototype and the low-impact requirements that characterize most industrial scale systems. We collected KPIs from all the tiers characterizing the system, by relying on already available services, when possible, and on ad-hoc build monitors otherwise.  We collected only KPIs that can be monitored with no functional impact and negligible performance overhead. As expected, \approach did not impose any limitation on the set of collected and processed KPIs, and we expect this to be valid in general. \begin{review} Studying the impact of noisy and redundant KPIs on the performance of PreMiSE may further improve the technique.
\end{review}

At the application tier, \approach collects both standard SNMP KPIs, such as communication latency between virtual machines, and Clearwater specific KPIs, such as the number of rejected IP-voice calls. 
At the IaaS tier, \approach collects KPIs about the cloud resource usage, such as the rate of read and write operations executed by OpenStack.
At the operating system tier, \approach collects KPIs about the usage of the physical resources, such as consumption of computational, storage, and network resources. In our evaluation, we used a sampling rate of 60 seconds.

\approach elaborates KPIs from both simple and aggregated metrics, that is, metrics that can be sampled directly, such as CPU usage, and metrics derived from multiple simple metrics, for example the call success rate, which can be derived from the number of total and successful calls, respectively.
The \emph{KPI Monitor} sends the data collected at each node to the predictor node that runs \approach on a Red Hat Enterprise Linux Server release 6.3 with Intel(R) Core (TM) 2 Quad Q9650 processor at 3GHz frequency and 16GB RAM. 

We implemented the \emph{Baseline Model Learner} and the \emph{Anomaly Detector}
on release 1.3 of OA-PI~\cite{IBM:SmartCloudAnalyticsPI:2015}, a state-of-the-art tool that computes the correlation between pairs of KPIs.
OA-PI  detects anomalies by implementing the following anomaly detection criteria: normal baseline variance, normal-to-flat variation, variance reduction, Granger causality, unexpected level, out-of-range values, rare values\footnote{\url{https://www.ibm.com/support/knowledgecenter/SSJQQ3_1.3.3/com.ibm.scapi.doc/intro/r_oapi_adminguide_algorithms.html}},
and issues alarms after revealing anomalies in few consecutive samples. 
\begin{change} OA-PI can analyze large volumes of data in real-time (the official IBM documentation show an example server configuration to  manage $1,000,000$ KPIs)\footnote{\url{https://www.ibm.com/support/knowledgecenter/en/SSJQQ3_1.3.3/com.ibm.scapi.doc/intro/c_tasp_intro_deploymentscenarios.html}}, and thus enables \approach to deal with the amount of KPIs that characterise large-scale distributed systems.
The OP-PI learning phase requires data from at least two weeks of normal operation behaviors, thus determining the \approach two weeks training interval. \end{change} 
%

We implemented the \emph{Signature Model Extractor} and the \emph{Failure Predictor} on top of
the Weka library~\cite{Weka:DataMining:2015}, a widely used open-source library that supports several classic machine learning algorithms.
We empirically compared the effectiveness of seven popular algorithms for solving classification problems when used to generate signatures:

\begin{itemize}
\item  a function-based \emph{Support Vector Machine (SVM)} algorithm that implements a sequential minimal optimization~\cite{Platt:SupportVectorMachine:SMO:1998},

\item a \emph{Bayesian Network (BN)} algorithm based on hill climbing~\cite{Cooper:BayesianNetwork:ML:1992},

\item a best-first \emph{Best-First Decision Tree (BFDT)} algorithm that builds a decision tree using a best-first search strategy~\cite{Friedman:BestFirstDT:ML:2000},

\item a \emph{Na\"ive Bayes (NB)} algorithm that implements a simple form of Bayesian network that assumes that the predictive attributes are conditionally independent, and that no hidden or latent attributes influence the prediction process~\cite{John:NaiveBayes:ML:1995},
 
\item a \emph{Decision Table (DT))} algorithm based on a decision table format that may include multiple exact condition matches for a data item, computing the result by a majority vote~\cite{Kohavi:DecisionTable:ML:1995},

\item a \emph{Logistic Model Tree (LMT)} algorithm that combines linear logistic regression and tree induction~\cite{Niels:LMT:ML:2005},

\item a \emph{Hidden Na\"ive Bayes (HNB)} algorithm that uses the mutual information attribute weighted method to weight one-dependence estimators~\cite{Zhang:HNB:ML:2005}.
\end{itemize}

As discussed in Section~\ref{sec:results} and illustrated in Table~\ref{tab:multiClassPredictions}, the results of our evaluation do not indicate major differences among the considered algorithms, with a slightly better performance of Logistic Model Trees that we adopt in the experiments.
 
\subsection{Fault Seeding}
\label{sec:faultSeeding}


In this section, we discuss the methodology that we followed to seed faults in the testbed. 
Fault seeding consists of introducing faults in a system to reproduce the effects of real faults, and is a common approach to evaluate the dependability of systems and study the effectiveness of fault-tolerance mechanisms~\cite{Bennett:ChaosMonkey:CloudDependabilityTest:2015,Sharma:CloudPD:DSN:2013} in test or production environments~\cite{Sauvanaud:Anomaly:ISSRE:2016,Blohowiak:FaultInjection:ISSRE:2016}. 

Since we use a cloud-based system to evaluate \approachNoSpace, we identify a set of faults that are well representative of the problems that affect cloud-based applications.
We analyze a set of issue reports\footnote{We conducted the analysis in July 2014 selecting the most recent issue reports at the time of the inspection.} of some relevant cloud projects to determine the most relevant fault types that threat Cloud applications. 
We analyze a total of 106 issue reports, 18 about KVM\footnote{https://bugzilla.kernel.org/buglist.cgi?component=kvm}, 62 about OpenStack\footnote{https://bugs.launchpad.net/openstack}, 19 about CloudFoundry\footnote{https://www.pivotaltracker.com/n/projects/956238}, and 7 about Amazon\footnote{http://aws.amazon.com}, and we informally assess the results with our industrial partners that operate in the telecommunication infrastructure domain.

\begin{figure}[!htbp] \centering
\includegraphics[width=10cm]{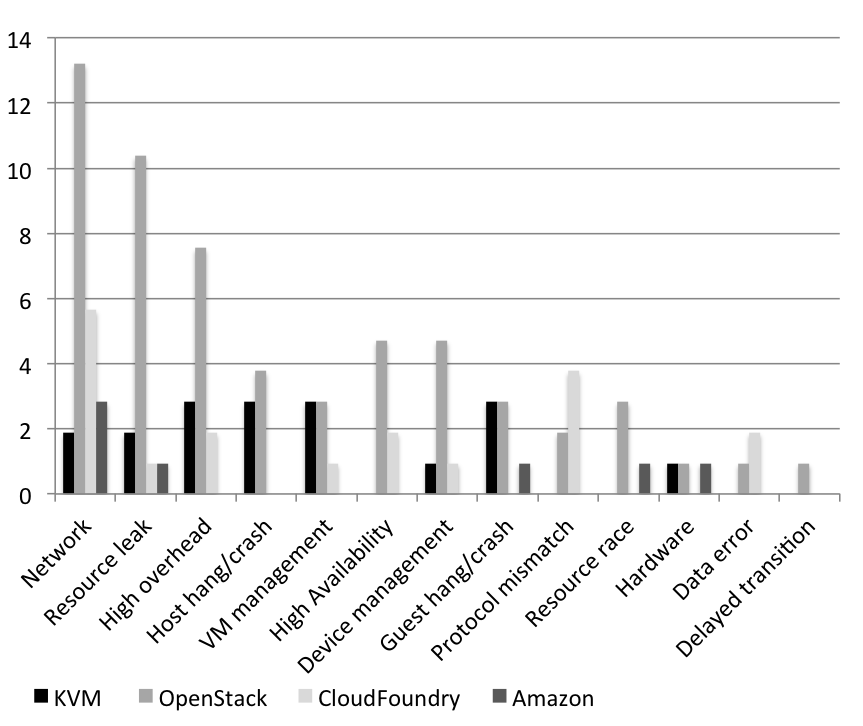}
\caption{Occurrences of categories of faults in the analyzed repositories}
\label{fig:failures}
\end{figure}

We classify the analyzed faults in thirteen main categories. Figure~\ref{fig:failures} plots the percentage of faults per category in decreasing order of occurrence for the analyzed fault repositories. 
The figure indicates a gap between the three more frequent and the other categories of faults, and we thus experimented with the three most frequent categories: \textit{Network}, \textit{Resource leaks} and \textit{High overhead} faults.
\textit{Network} faults consist of networking issues that typically affect the network and transport layers, such as packet loss problems.
\textit{Resource leaks} occur when resources that should be available for executing the system are not obtainable, for instance because a faulty process does not release memory when not needed any more.
\textit{High overhead} faults occur when a system component cannot
meet its overall objectives due to inadequate performance, for instance because of poorly implemented APIs or resource-intensive activities.  


Based on the results of this analysis, we evaluate \approach with injected faults of six types, that characterize the three top ranked categories of faults in Figure~\ref{fig:failures}:  \emph{Network faults} that depend on \emph{Packet Loss} due to \emph{Hardware} and \emph{Excessive workload} conditions, increased \emph{Packet Latency} due to network delay and \emph{Packet corruption} due to errors in packet transmission and reception, \emph{Resource leak faults} that depend on \emph{Memory Leaks}, and \emph{High overhead faults} that depend on \emph{CPU Hogs}. In details,
\begin{inparaenum}[(i)]
\item A packet loss due to hardware conditions drops a fraction of the network packets, and simulates the degradation of the cloud network; 
\item  A packet loss due to excessive workload conditions corresponds to an extremely intensive workload, and causes an intensive packet loss;
\item An increased packet latency and 
\item corruption due to channel noise, routing anomalies or path failures, simulates degraded packet delivery performances;
\item A memory leak fault periodically allocates some memory without releasing it, simulates a common software bug, which severely threaten the dependability of cloud systems;
\item A CPU Hog fault executes some CPU intensive processes that consume most of the CPU time and cause poor system performance.
\end{inparaenum}

We limited our investigation to the most relevant categories of faults to control the size of the experiment, which already involves an extremely large number of executions. 
The results that we discuss in Section~\ref{sec:results} demonstrate the effectiveness of \approach across all the faults considered in the experiments.  We expect comparable results for other fault categories with the same characteristics of the considered ones, namely faults that lead to the degradation of some KPI values over time before a failure.  This is the case of most of the fault categories of Figure~\ref{fig:failures}, with the exception of host and guest crashes, which may sometime occur suddenly and without an observable degradation of KPI values over time.  Confirming this hypothesis and thus extending the results to a broader range of fault categories would require additional experiments.

We inject packet loss, packet latency, packet corruption, memory leaks and CPU Hogs faults into both the host (Openstack) and guest (Clearwater) layers, and excessive workload faults by altering the nature of the workload, following the approaches proposed in previous studies on dependability of cloud systems~\cite{Bennett:ChaosMonkey:CloudDependabilityTest:2015} and on problem determination in dynamic clouds~\cite{Sharma:CloudPD:DSN:2013}. 


We study a wide range of situations by injecting faults according to three activation patterns:
\begin{description}
  \item[Constant:]  the fault is triggered with a same frequency over time.
  \item[Exponential:]  the fault is activated with a frequency that increases exponentially, resulting in a shorter time to failure.
  \item[Random:]  the fault is activated randomly over time.
\end{description}

Overall, we seeded 12 faults in different hosts and VMs. Each fault is
characterized by a fault type and an activation pattern.

\subsection{Workload Characteristics}
\label{sec:workload}

\begin{review}In practice, it is hard to know if a set of executions is general enough. In our specific settings, we define\end{review} the workload used in the experimental evaluation to replicate the shape of real SIP traffic as experienced by our industrial partners in the telecommunication domain. 
We carefully tuned the peak of the workflow to use as much as 80\% of CPU and memory.
We generate the SIP traffic with the SIPp traffic generator~\cite{Gayraud:SIPpTrafficGenerator:2015}, which is an open source initiative from Hewlett-Packard (HP) and is the de facto standard for SIP performance benchmarking.



SIPp can simulate the generation of multiple calls using a single machine. The generated calls follow user-defined scenarios that include the exact definition of both the SIP dialog and the structure of the individual SIP messages. In our evaluation, we use the main SIP call flow dialogs as documented in Clearwater\footnote{http://www.projectclearwater.org/technical/call-flows/}.

\begin{figure}[!htbp] \centering 
\includegraphics[width=12cm]{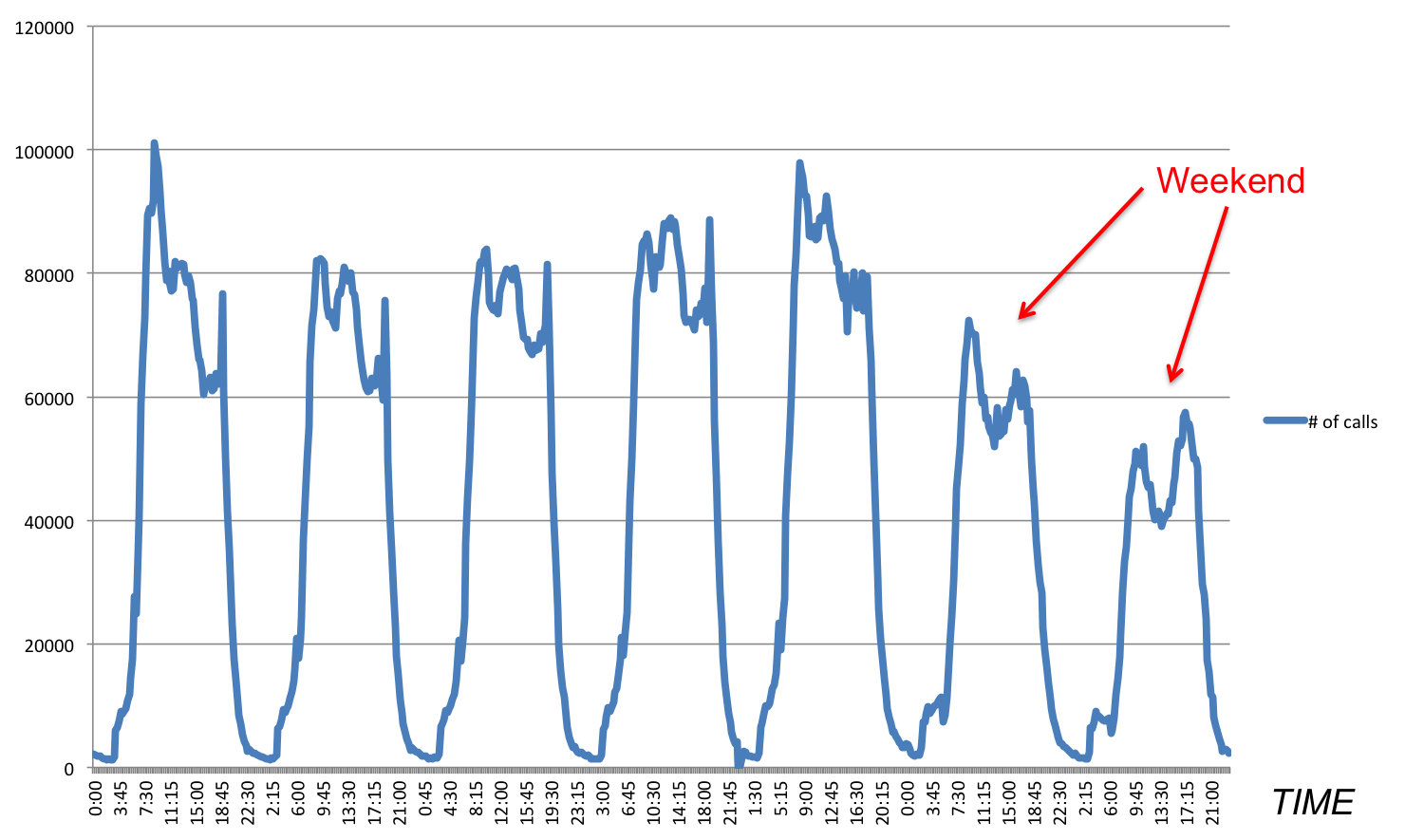}
\caption{Plot with calls per second generated by our workload over a week}
\label{fig:weeklyWorkload}
\end{figure}

\begin{figure}[!htbp] \centering
\includegraphics[width=12cm]{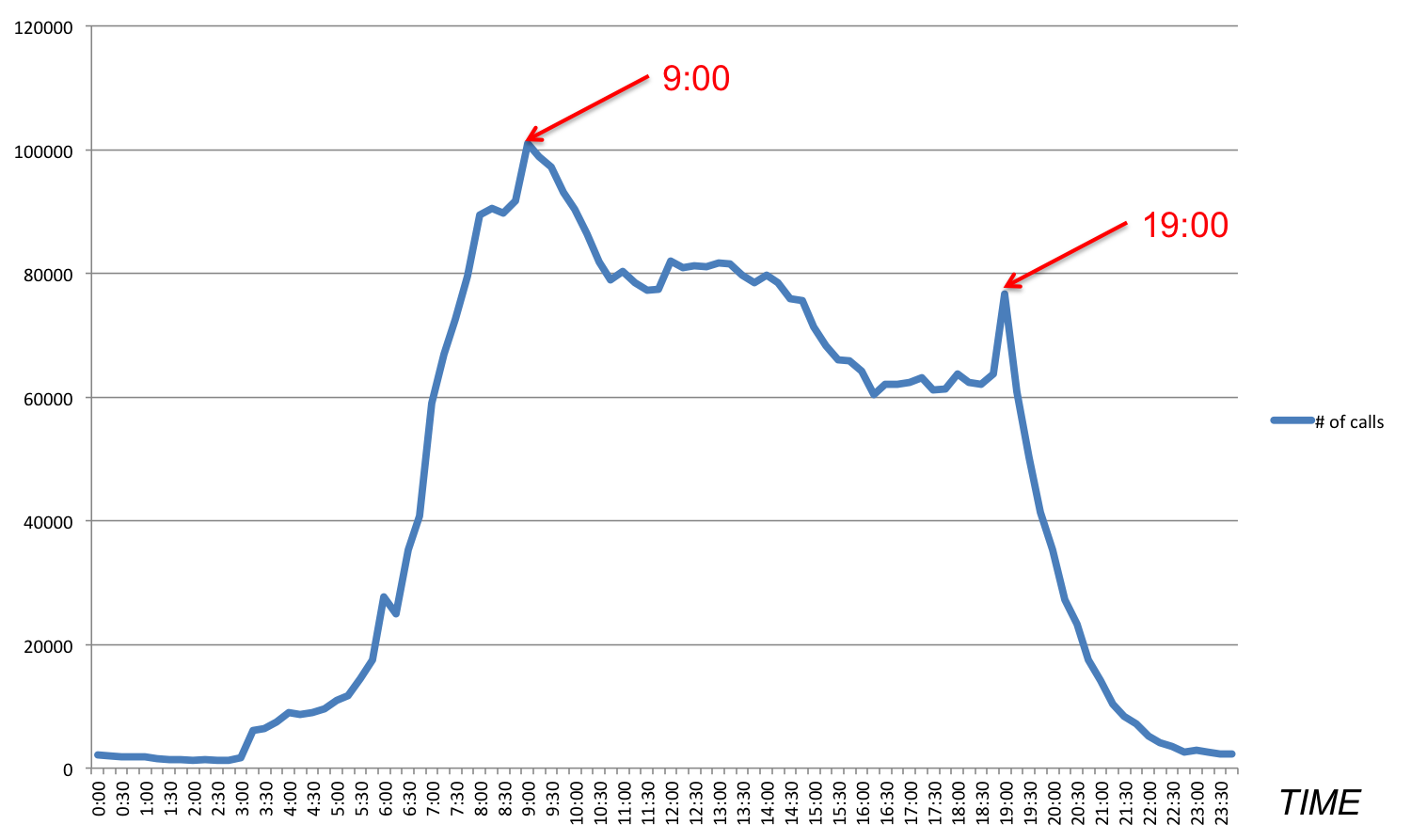}
\caption{Plot with calls per second generated by our workload over a day}
\label{fig:dailyWorkload}
\end{figure}

Our workload includes a certain degree of randomness, and generates new calls based on a call rate that changes according to calendar patterns. In particular, we consider two workload patterns:

\begin{description}
  \item[Daily variations] The system is busier on certain days of the week. In particular, we consider a higher traffic in working days (Monday through Friday) and lower traffic in the weekend days (Saturday and Sunday). 
  Figure~\ref{fig:weeklyWorkload} graphically illustrates the structure of our workload over a period of a week. 
  \item[Hourly variations] To resemble daily usage patterns, our workload is lighter during the night and heavier during the day with two peaks at 9am and 7pm, as graphically illustrated in Figure~\ref{fig:dailyWorkload}.
\end{description}

\begin{review}In our empirical evaluation, we obtained good results already with the workload that we designed, without the need of introducing extensive variability in the normal executions used for training. This is probably a positive side effect of the usage of anomaly detection and failure prediction in a pipeline. In fact, the failure predictor component can compensate the noise and false positives produced by anomaly detector.\end{review}

\subsection{Evaluation Measures}
\label{sec:qualityMetrics}

We addressed the research questions RQ1, RQ2 and RQ5 by using 10-fold cross-validation~\cite{Witten:dataMining:2011}. \approach analyzes time series data, and collects anomalous KPIs every 5 minutes, to comply with the requirements of IBM OA-PI~\cite{IBM:SmartCloudAnalyticsPI:2015}, the time series analyzer used in \approach.  Signature-base analysis does not consider the order in which the sliding windows are arranged, thus we collected the \emph{samples} necessary to apply 10-fold cross validation during the execution of our workload with sliding windows of length $l$. Since each run lasts 120 minutes and 
the size of the interval in the sliding window is 5 minutes, each workload execution produces $(120-l)/5$ samples that can be used for prediction. In the evaluation, we first studied the impact of $l$ on the results (RQ1), and then used the best value in our contest for the other experiments.  

Overall we collected samples from a total of 648 runs, which include 24 passing executions and 24 failing executions for each type of failure. 
A failing execution is characterized by a fault of a given type injected in a given resource with a given activation pattern.  As discussed in Section~\ref{sec:faultSeeding}, we injected faults of six different types (packet loss, excessive workload, packet latency, packet corruption, memory leak and cpu hog) following three activation patterns
(constant, exponential and random). For all but excessive workload, we injected
faults on five different target resources (the Bono, Sprout, and Homestead virtual machines in Clearwater and two compute nodes in OpenStack), resulting in $5\times 3\times 5=75$ failure cases. For excessive workload, we  injected faults with three patterns with no specific target resource, since excessive workload faults target the system and not a specific resource. We thus obtained $75+3=78$  failure cases. To avoid biases due to the fault injection pattern, we repeated every experiment 8 times, thus obtaining 624 failing executions for the evaluation. The extensive investigation of the different fault types, activation patterns, and affected resources made the set of executions available for the experiment unbalanced between passing executions (24 cases) and failing executions (624 cases). 


Since we use $l=90$ for RQ2 and RQ5, we obtained a total of 4,782 samples collected from both passing and failing executions. The number of samples available for RQ1 is higher because we tried different values for $l$. To apply 10-fold cross-validation, we split the set of samples into 10 sets of equals size, using nine of them to learn the prediction model and the remaining set to compute the quality of the model. The \approach failure prediction algorithm does not consider the order of the samples in time, since it classifies each sample independently from the others.


We evaluated the quality of a prediction model using the standard measures that are used to define contingency tables and that cover the four possible outcomes of failure prediction (see Table~\ref{tab:contingencyTable}). We also measured the following derived metrics:

\begin{description}
  \item[Precision:] the ratio of correctly predicted failures over all predicted failures. This measure can be used to assess the rate of false alarms, and thus the rate of unnecessary reactions that might be triggered by the failure predictor.
  \item[Recall:] the ratio of correctly predicted failures over actual failures. This measure can be used to assess the percentage of failures that can be predicted with the failure predictor.
  \item[F-Measure:] the uniformly weighted harmonic mean of precision and recall. This measure captures with a single number the tradeoff between precision and recall. 
  \item[Accuracy:] the ratio of correct predictions over the total number of predictions. The accuracy provides a quantitative measure of the capability to predict both failures and correct executions.
   \item[FPR (False Positive Rate):] the ratio of incorrectly predicted failures to the number of all correct executions. The FPR provides a measure of the false warning frequency.
\end{description}

Table~\ref{tab:contingencyTableMetrics} summarizes the derived metrics that we used by presenting their mathematical formulas and meanings.

\begin{table}[h]
\caption {Contingency table}
\label{tab:contingencyTable}
\vspace{-0.4cm}
\begin{center}
\begin{tabular}{cc|c|c|}
\cline{3-4}
        &             &
                           \multicolumn{2}{c|}{\textbf{Predicted}}
                        \\ \cline{3-4} &             & Failure    & Not-Failure \\ \hline
\multicolumn{1}{|c|}{\multirow{2}{*}{\textbf{Actual}}} & Failure     &
\begin{tabular}[c]{@{}c@{}}True Positive (TP)\\ \textit{(correct
warning)}\end{tabular}  & \begin{tabular}[c]{@{}c@{}}False Negative (FN)\\
\textit{(missed warning)}\end{tabular}       \\ \cline{2-4}
\multicolumn{1}{|c|}{} & Not-failure & \begin{tabular}[c]{@{}c@{}}False Positive (FP)\\
\textit{(false warning)}\end{tabular} & \begin{tabular}[c]{@{}c@{}}True
Negative (TN)\\
\textit{(correct no-warning)}\end{tabular} \\ \hline
\end{tabular}
\end{center}
\end{table}

\begin{table}[h]
\caption {Selected metrics obtained from the contingency table}
\label{tab:contingencyTableMetrics}
\begin{center}
\begin{tabular}{|c|c|c|}
\hline
Metric    & Formula & Meaning                                                                                                          \\ \hline
 &   & How many predicted \\ Precision& $\frac{TP}{(TP+FP)}$ & failures
 are actual \\
 & &
failures? \\ \hline     &   & How many actual
\\ Recall&$\frac{TP}{(TP+FN)}$  & failures are correctly \\ & &
predicted as failures? \\ \hline
 &   & Harmonic mean of
\\ F-measure&$2*\frac{(Precision*Recall)}{(Precision+Recall)}$  & $Precision$
and $Recall$
\\
& &  \\ \hline
 &   & How many predictions
\\ Accuracy&$\frac{(TP+TN)}{(TP+TN+FP+FN)}$  & are correct?
\\
& &  \\ \hline
 &   & How many correct 
\\ FPR&$\frac{(FP)}{(TN+FP)}$  &  executions are 
\\
& &  predicted as failures? \\ \hline
\end{tabular}
\end{center}
\vspace{-0.5cm}
\end{table}


We addressed the research question RQ3 by computing the \emph{percentage of samples that \approach correctly classifies as failure-free} given a set of samples collected by running workflows that differ significantly from the workflow used during the training phase. To this end, we designed two new workflows: \emph{random40} and \emph{random100}. 
The \emph{random40} workflow behaves like the training
workflow with a uniformly random deviation between 0\% to 40\%, while the
\emph{random100} workflow behaves with a deviation of up to 100\%.

\begin{figure}
\begin{center}
  \includegraphics[width=9.0cm]{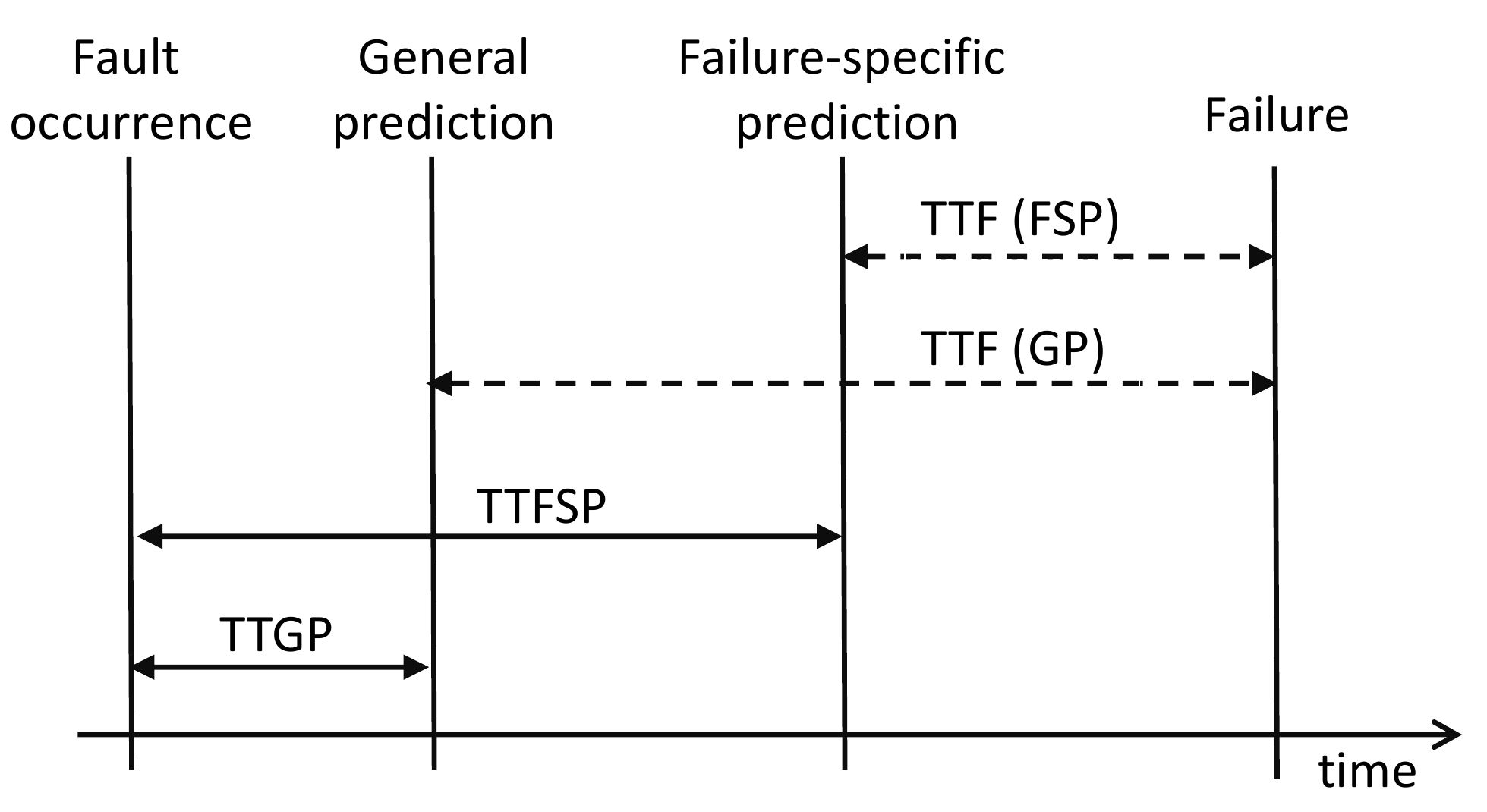}
\caption{Prediction time measures}
\label{fig:predictionTime}
\end{center}
\vspace{-0.7cm}
\end{figure}

We addressed the research question RQ4 by computing \emph{the time needed to
generate a prediction} and \emph{the time between a failure prediction and its
occurrence} from a total of 18 faulty runs lasting up to twelve hours. The former time
measures the capability of \approach to identify and report erroneous behaviors. 
The latter time estimates how early \approach can predict failure occurrences. As shown in Figure~\ref{fig:predictionTime}, we define four specific measures:

\begin{description}
  \item[Time-To-General-Prediction (\emph{TTGP}):] the distance between the time a fault is active for the first time and the time \approach produces a general prediction,
    \item[Time-To-Failure-Specific-Prediction (\emph{TTFSP}):] the distance between the time a fault is active for the first time and the time \approach predicts a specific failure type,
    \item[Time-To-Failure for General Prediction  (\emph{TTF(GP)}):] the distance between the time \approach predicts a general failure and the time the failure happens,
    \item[Time-To-Failure for Failure-Specific Prediction  (\emph{TTF(FSP)}):] the distance between the time \approach predicts a specific failure type and the time the system fails,
\end{description}

where the \emph{Fault occurrence} is the time the seeded fault becomes active in the system, the \emph{General prediction} is the first time \approach signals the presence of a failure without indicating the fault yet, that is, it identifies an anomaly with an empty fault and resource, the \emph{Failure-specific prediction} is the first time \approach indicates also the fault type and the faulty resource, the \emph{Failure} is the time the delivered service deviated from the system function. Failures depend on the seeded faults. In our case, failures manifest either as system crashes or as success rate dropping below 60\%, as indicated in Section~\ref{sec:results} when discussing RQ4.



To answer RQ6, we measured the resource consumption as
\begin{inparaenum}[(i)] 
\item the percentage of \emph{CPU} used by the monitoring activities,
\item the amount of \emph{memory} used by the monitoring activities,
\item the amount of \emph{bytes read/written} per second by the monitoring activities, and
\item the packets received/sent per second over the \emph{network} interfaces by the monitoring activities.
\end{inparaenum}

\section{Experimental Results}
\label{sec:results}

In this section we discuss the results of the experiments that we executed to
answer the research questions that we introduced in the former section.



\begin{figure} \centering \includegraphics[width=8.7cm]{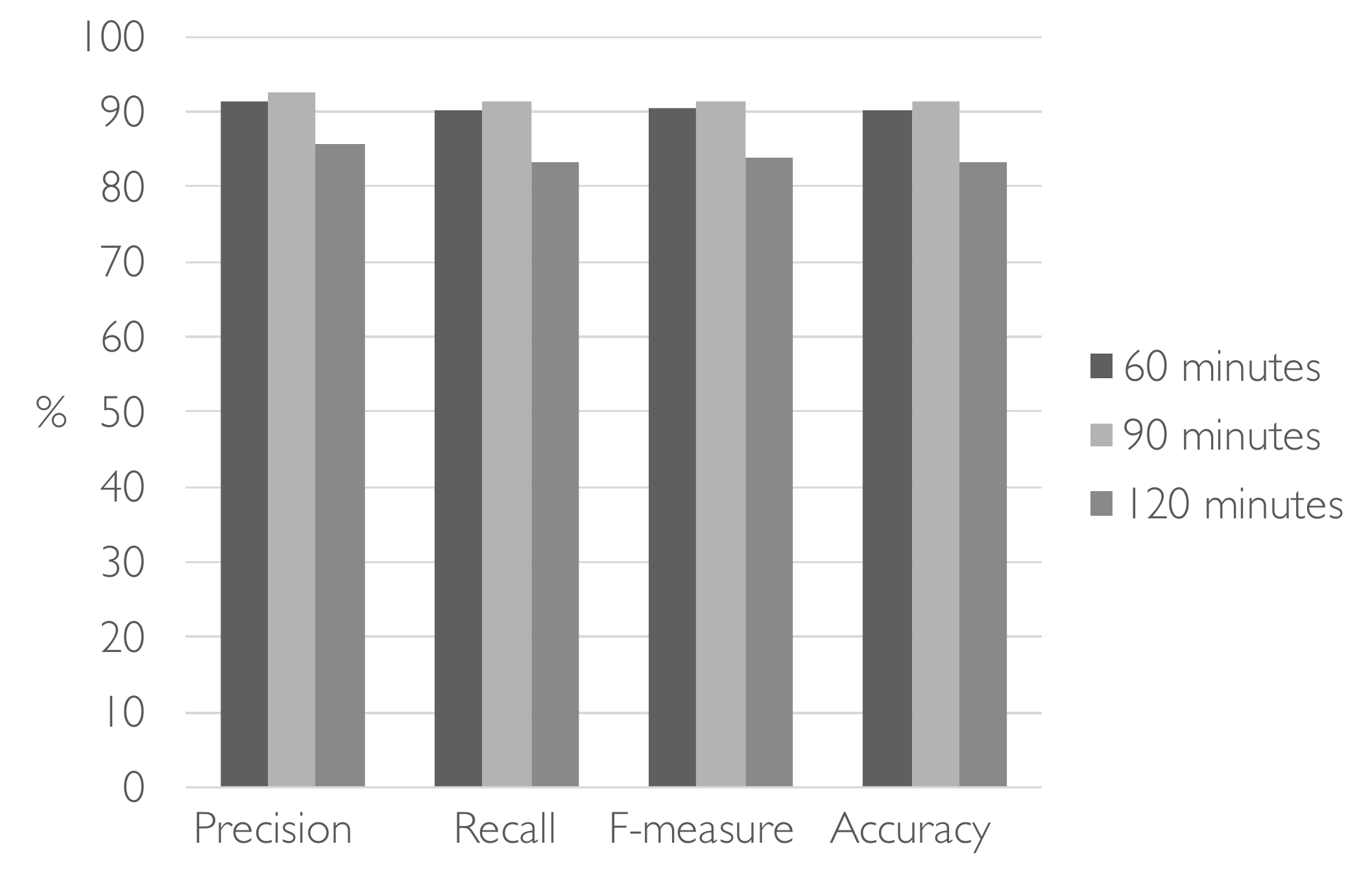}
\caption{Average effectiveness of failure prediction approaches with different sliding window sizes}
\label{fig:observationWindow}
\vspace{-0.4cm}
\end{figure}

\begin{figure} \centering \includegraphics[width=8cm]{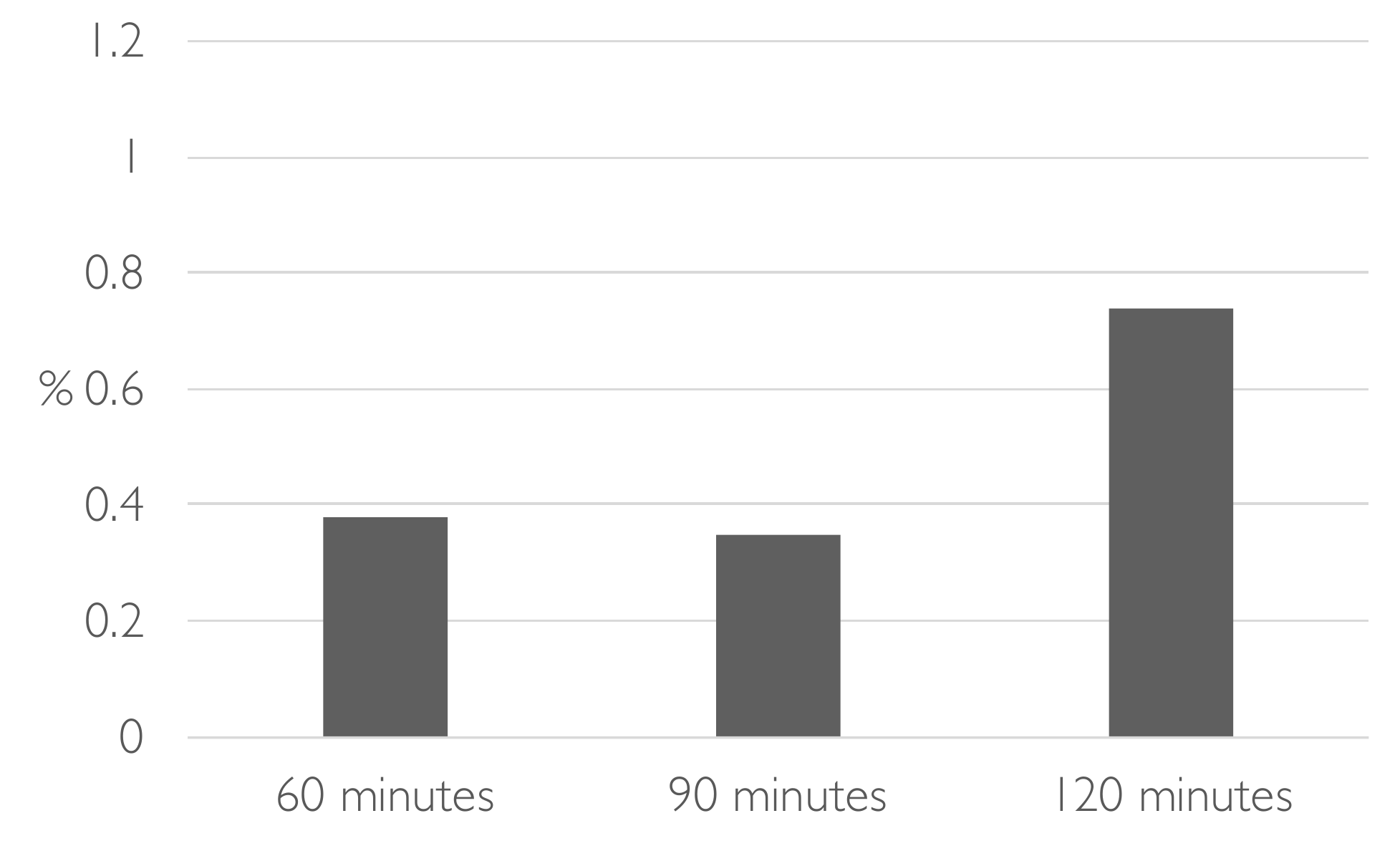}
\caption{Average false positive rate for different sliding window sizes}
\label{fig:fprObservationWindow}
\vspace{-0.3cm}
\end{figure}

\begin{table}
\caption {Comparative evaluation of the effectiveness of \approach prediction and localization with the different algorithms for generating signatures}
\label{tab:multiClassPredictions}
\vspace{-0.2cm}
\begin{center}
\begin{tabular}{|l||r|r|r|r|r|}
 \hline
\emph{Model}  & \emph{Precision} & \emph{Recall} & \emph{F-measure} &
\emph{Accuracy} & \emph{FPR}
\\\hline
\hline 
\emph{BN} & 	84.906 & 	82.438 & 	82.430 & 	82.438 & 	0.685 \\ \hline
\emph{BFDT} & 	97.746 & 	97.719 & 	97.726 & 	97.719 & 	0.093 \\ \hline
\emph{NB} & 	83.931 & 	80.925 & 	80.786 & 	80.925 & 	0.745 \\ \hline
\emph{SVM} & 	98.632 & 	98.632 & 	98.632 & 	98.632 & 	0.057 \\ \hline
\emph{DT} & 	92.650 & 	88.555 & 	89.943 & 	88.555 & 	0.470 \\ \hline
\emph{LMT} & 	98.798 & 	98.797 & 	98.797 & 	98.797 & 	0.050 \\ \hline
\emph{HNB} & 	92.307 & 	91.831 & 	91.765 & 	91.831 & 	0.325 \\ \hline
\end{tabular}
\end{center}
\vspace{-0.2cm}
\end{table}

\subsubsection*{RQ1: Sliding window size}
\label{sec:slidingWindowResults}

\approach builds the prediction models and analyses anomalies referring to time sliding windows of fixed size. 
The sliding windows should be big enough to contain a sufficient amount of information to predict failures and small enough to be practical and sensitive to failure symptoms.

With this first set of experiments, we investigate the impact of the window size on the effectiveness of the prediction.  
We experimented with the seven algorithms described in Section~\ref{sec:implementationDetails}, each with sliding windows of size 60, 90 and 120 minutes to study the impact of the window size, and to chose the size for the next experiments. We built a total of 27 prediction models.
We executed the prototype tool with the different prediction models both with and without seeded faults, for a total of 24 execution for 27 configurations, 26 of which corresponding to configurations each seeded with a different fault, and one no-faulty configuration, for a total of 648 executions. The configurations correspond to the raw of Table~\ref{tab:LMTperformance} that we discuss later in this section.

%

Figure~\ref{fig:observationWindow} compares the average precision, recall, F-measure and accuracy over all the experiments. 
These results indicate that the window size has a moderate impact on the predictions, and that a window size of 90 minutes reaches the best prediction effectiveness among the experimented sizes. 
Figure~\ref{fig:fprObservationWindow} shows the average false positive rates for the different window sizes, and confirms the choice of a window of 90 minutes as the optimal choice among the evaluated sizes. The results collected for the individual algorithms are consistent with the average ones. 
In all the remaining experiments, we use 90-minutes windows. 

\subsubsection*{RQ2: Predicting Failures and locating faults}


We evaluated the effectiveness of \approach as the ability of predicting incoming failures and identifying the kind and location of the related faults.

Table~\ref{tab:multiClassPredictions} shows the precision, recall, F-measure, accuracy and False Positive Rate (FPR) of failure prediction and fault localization for \approach with the prediction algorithms presented at the end of Section~\ref{sec:implementationDetails}.
The table indicates that \approach performs well with all the algorithms, with slightly better indicators for \emph{LMTs (Logistic Model Trees)} that we select for the remaining experiments.

Table~\ref{tab:LMTperformance} shows the effectiveness of \approach with LMT for
the different fault types and locations.  The metrics were calculated on a window basis as you need to make a forecast about each window. This means that windows that belong to both failed and correct executions are taken into account. The results in the table indicate that
the approach is extremely accurate: \approach suffered from only~74 false
predictions out of~4,782 window samples.
\approach can quickly complete the offline training phase. To learn the \emph{baseline model}, the data collected from two weeks of execution
required less than 90 minutes of processing time. When the training phase runs in parallel to the data collection process, it completes almost immediately after the data collection process has finished. The \emph{signature model extractor} has taken less than 15 minutes to be learnt using the anomalies from two weeks.


 
 \begin{table}
 \begin{center}
 \begin{scriptsize}
 \caption {Effectiveness of the LogicModel tree (LMT) failure prediction  algorithm for fault type and location}
 \label{tab:LMTperformance}
 \begin{tabular}{|c|c|c|c|c|c|}
 \hline
 \begin{tabular}[c]{@{}c@{}}Fault type (Location) \end{tabular} & Precision &
 Recall & F-Measure & Accuracy & FPR \\ \hline
\begin{tabular}[c]{@{}c@{}}CPU hog (Bono) \end{tabular}   & 	100\% & 	93.529\% & 	96.657\% & 	0.998\% & 	0\%                  \\ \hline
\begin{tabular}[c]{@{}c@{}}CPU hog (Sprout) \end{tabular}   & 	100\% & 	97.059\% & 	98.507\% & 	0.999\% & 	0\%          \\ \hline
\begin{tabular}[c]{@{}c@{}}CPU hog (Homestead) \end{tabular}   &	100\% & 	97.041\% & 	98.498\% & 	0.999\% & 	0\%          \\ \hline
\begin{tabular}[c]{@{}c@{}}CPU hog (Compute \#5) \end{tabular}   & 	93.820\% & 	98.817\% & 	96.254\% & 	0.997\% & 	0.236\%          \\ \hline
\begin{tabular}[c]{@{}c@{}}CPU hog (Compute \#7) \end{tabular}   & 	96.875\% & 	91.716\% & 	94.225\% & 	0.996\% & 	0.107\%          \\ \hline
\begin{tabular}[c]{@{}c@{}}Excessive workload \end{tabular}   & 	100\% & 	100\% & 	100\% & 	1.000\% & 	0\%                  \\ \hline
\begin{tabular}[c]{@{}c@{}}Memory leak (Bono) \end{tabular}   & 	100\% & 	98.810\% & 	99.401\% & 	1.000\% & 	0\%          \\ \hline
\begin{tabular}[c]{@{}c@{}}Memory leak (Sprout) \end{tabular}   &	100\% & 	95.833\% & 	97.872\% & 	0.999\% & 	0\%          \\ \hline
\begin{tabular}[c]{@{}c@{}}Memory leak (Homestead) \end{tabular}   &	100\% & 	96.429\% & 	98.182\% & 	0.999\% & 	0\%          \\ \hline
\begin{tabular}[c]{@{}c@{}}Memory leak (Compute \#5) \end{tabular}   &	76.119\% & 	91.071\% & 	82.927\% & 	0.987\% & 	1.031\%          \\ \hline
\begin{tabular}[c]{@{}c@{}}Memory leak (Compute \#7) \end{tabular}   &	93.333\% & 	75.000\% & 	83.168\% & 	0.989\% & 	0.193\%          \\ \hline
\begin{tabular}[c]{@{}c@{}}Packet corruption (Bono) \end{tabular}   &	85.973\% & 	99.476\% & 	92.233\% & 	0.993\% & 	0.669\%          \\ \hline
\begin{tabular}[c]{@{}c@{}}Packet corruption (Sprout) \end{tabular}   &	87.558\% & 	99.476\% & 	93.137\% & 	0.994\% & 	0.583\%          \\ \hline
\begin{tabular}[c]{@{}c@{}}Packet corruption (Homestead) \end{tabular}   &	99.429\% & 	91.579\% & 	95.342\% & 	0.996\% & 	0.022\%  \\ \hline
\begin{tabular}[c]{@{}c@{}}Packet corruption (Compute \#5) \end{tabular}   &	100\% & 	100\% & 	100\% & 	1.000\% & 	0\%          \\ \hline
\begin{tabular}[c]{@{}c@{}}Packet corruption (Compute \#7) \end{tabular}   &	100\% & 	100\% & 	100\% & 	1.000\% & 	0\%          \\ \hline
\begin{tabular}[c]{@{}c@{}}Packet latency (Bono) \end{tabular}   &	96.000\% & 	100\% & 	97.959\% & 	0.998\% & 	0.173\%          \\ \hline
\begin{tabular}[c]{@{}c@{}}Packet latency (Sprout) \end{tabular}   &	76.777\% & 	84.375\% & 	80.397\% & 	0.984\% & 	1.058\%          \\ \hline
\begin{tabular}[c]{@{}c@{}}Packet latency (Homestead) \end{tabular}   &	72.028\% & 	53.646\% & 	61.493\% & 	0.973\% & 	0.864\%  \\ \hline
\begin{tabular}[c]{@{}c@{}}Packet latency (Compute \#5) \end{tabular}   &	82.857\% & 	75.521\% & 	79.019\% & 	0.984\% & 	0.648\%          \\ \hline
\begin{tabular}[c]{@{}c@{}}Packet latency (Compute \#7) \end{tabular}   &	62.069\% & 	75.000\% & 	67.925\% & 	0.972\% & 	1.900\%          \\ \hline
\begin{tabular}[c]{@{}c@{}}Packet loss (Bono) \end{tabular}   &	100\% & 	73.837\% & 	84.950\% & 	0.991\% & 	0\%          \\ \hline
\begin{tabular}[c]{@{}c@{}}Packet loss (Sprout) \end{tabular}   &	99.429\% & 	100\% & 	99.713\% & 	1.000\% & 	0.022\%          \\ \hline
\begin{tabular}[c]{@{}c@{}}Packet loss (Homestead) \end{tabular}   &	94.152\% & 	95.833\% & 	94.985\% & 	0.996\% & 	0.215\%          \\ \hline
\begin{tabular}[c]{@{}c@{}}Packet loss (Compute \#5) \end{tabular}   &	100\% & 	100\% & 	100\% & 	1.000\% & 	0\%          \\ \hline
\begin{tabular}[c]{@{}c@{}}Packet loss (Compute \#7) \end{tabular}   &	82.266\% & 	99.405\% & 	90.027\% & 	0.992\% & 	0.773\%          \\ \hline
\begin{tabular}[c]{@{}c@{}}Correct execution \end{tabular}   &	100\% & 	100\% & 	100\% & 	1.000\% & 	0\%          \\ \hline
 \end{tabular}
 \end{scriptsize}
 \end{center}
 \vspace{-0.5cm}
 \end{table}
  

\subsubsection*{RQ3: Detecting Legal Executions}
\label{sec:legalExec}

While the workload conditions do not alter the failure detection and fault localization capabilities, they may impact on the false positive rate in the absence of faults.  Thus experimented with different types of workloads in the absence of faults:
workload \emph{random40} that differs from the workload used in the training phase for 40\% of the cases, and \emph{random100} that differs completely from the workload used in the training phase. 

We generated 72 samples for \emph{random40} and \emph{random100} by running each workload for 2 hours, producing a total of 144 samples. 
\approach has been able to correctly classify all the samples as belonging to failure-free executions. 
Jointly with the results 
discussed for RQ2, we can say that  \approach shows a very low number of false positives, even if we are analyzing data from normal executions with workloads completely different from those used in the training phase.

\subsubsection*{RQ4: Prediction Earliness}
\label{sec:predictionTimeResults}

We evaluated the earliness of the prediction as the Time-To-General-Prediction (\emph{TTGP}), the Time-To-Failure-Specific-Prediction (\emph{TTFSP}), the Time-To-Failure for General Prediction  (\emph{TTF(GP)}) and the Time-To-Failure for Failure-Specific Prediction (\emph{TTF(FSP)} illustrated in Figure~\ref{fig:predictionTime}.

In the experiments, failures correspond to either system crashes or drops in the successful SIP call rate below 60\% for 5 consecutive minutes. 
%
Table~\ref{tab:predictionTime} reports the results of the experiment. The columns \emph{from fault occurrence to failure prediction} show the time that \approach needed to predict a general (TTGP) and specific (TTFSP)  failure, respectively.
\approach has been able to produce a general failure prediction in some minutes: 5 minutes in the best case, less than 31 minutes for most of the faults, and 65 minutes in the worst case. 
Moreover, \approach has generated the failure specific prediction few minutes after the general prediction, with a worst case of 35 minutes from the general to the specific prediction. 
The readers should notice that we measure the time to prediction starting with the first activation of the seeded fault, which may not immediately lead to faulty symptoms. 

The columns  \emph{From failure prediction to failure} indicate that the failures are predicted well in advance, leaving time for a manual resolution of the problem.
\approach has detected both the general and failure specific predictions at least 48 minutes before the failure, which is usually sufficient for a manual intervention. These results are also valuable for the deployment of self-healing routines, which might be activated well in advance with respect to failures.


\approach predicts failure based on the analysis of \emph{OA-PI}, which works with sampling intervals of 5 minutes. Indeed \approach can effectively predict a failure with few anomalous samples. 
It could predict failures in a shorter time than 5 minutes with an anomaly detector that requires smaller sampling intervals.

\begin{figure}[ht!]
\begin{center}
  \includegraphics[width=12cm]{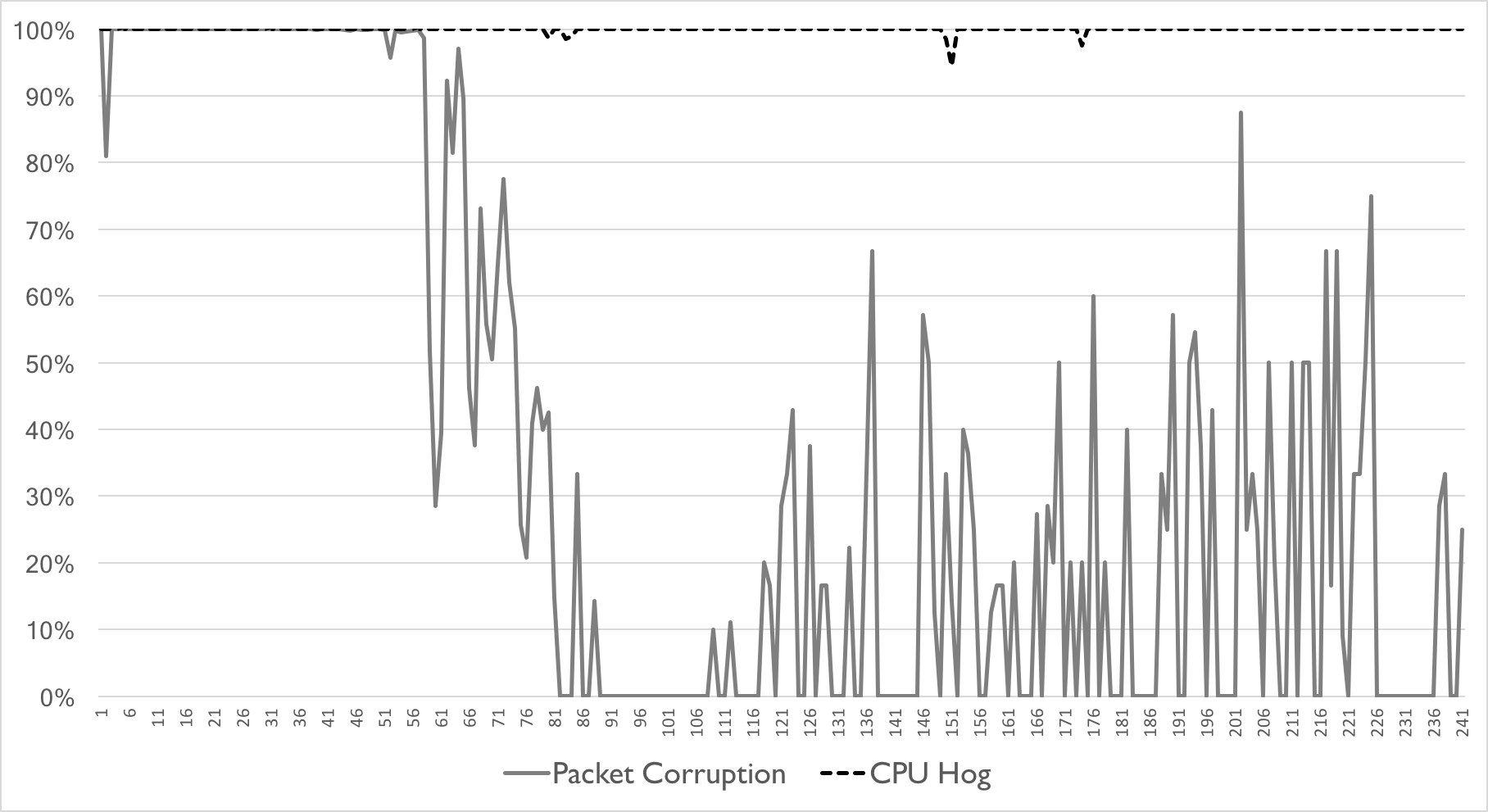}
\caption{Call success rate over time}
\label{fig:successrate}
\end{center}
\end{figure}

Faults of different type have very different impacts on the system, and can thus result in largely different patterns.
Figure~\ref{fig:successrate} exemplifies the different impact of faults of various types by plotting the percentage of successful calls in the experiments characterized by the longest and shortest Time-to-Failure, which correspond to CPU hog and packet corruption faults, respectively. 
Packet corruption faults have a gradual impact on the system, while the CPU hog faults do not cause failures in the first three hours of execution for the reported experiment.

Overall, \approach demonstrated to be able to effectively predict failures, including their type well in advance to the failure time for the four classes of problems that have been investigated.

\begin{table}
\caption {\approach prediction earliness for fault type and pattern}
\label{tab:predictionTime}
\begin{center}
\begin{tabular}{| l| | l | l |l |l|}
 \hline
  &   \multicolumn{2}{c|}{From fault occurrence} &   \multicolumn{2}{c|}{From failure prediction} \\ 
 &   \multicolumn{2}{c|}{to failure prediction} &   \multicolumn{2}{c|}{to failure}\\
 {\it \begin{tabular}[c]{@{}l@{}} Fault Type (Pattern)\end{tabular}} & \emph{TTGP} &
 \emph{TTFSP} &\emph{TTF 
 (GP)}
 & \emph{TTF 
 (FSP)}
\\
 
\hline
\hline 

{\it \begin{tabular}[c]{@{}l@{}}CPU hog (Random)\end{tabular}}
&65 mins& 80 mins & $>$12 hours & $>$12 hours
\\
\hline 
{\it \begin{tabular}[c]{@{}l@{}}CPU hog (Constant)\end{tabular}}
&45 mins& 60 mins & $>$12 hours & $>$12 hours
\\
\hline 
{\it \begin{tabular}[c]{@{}l@{}}CPU hog (Exponential)\end{tabular}}
&5 mins& 30 mins & $>$12 hours & $>$12 hours
\\
\hline 
{\it \begin{tabular}[c]{@{}l@{}}Excessive workload \\ (Random)\end{tabular}}
&35 mins& 50 mins & 192 mins & 177 mins
\\
\hline 
{\it \begin{tabular}[c]{@{}l@{}}Excessive workload \\ (Constant)\end{tabular}}
&40 mins & 55 mins &110 mins &95 mins
\\
\hline 
{\it \begin{tabular}[c]{@{}l@{}}Excessive workload \\ (Exponential)\end{tabular}}
&30 mins& 45 mins & 80 mins & 65 mins
\\
\hline 
{\it \begin{tabular}[c]{@{}l@{}}Memory leak (Random)\end{tabular}}
&5 mins& 20 mins & 55 mins & 40 mins
\\
\hline 
{\it \begin{tabular}[c]{@{}l@{}}Memory leak (Constant)\end{tabular}}
&5 mins& 20 mins & 56 mins & 41 mins
\\
\hline 
{\it \begin{tabular}[c]{@{}l@{}}Memory leak \\ (Exponential)\end{tabular}}
&5 mins& 20 mins & 56 mins & 41 mins
\\
\hline 
{\it \begin{tabular}[c]{@{}l@{}}Packet corruption \\ (Random)\end{tabular}}
&30 mins &60 mins &121 mins &91 mins
\\
\hline 
{\it \begin{tabular}[c]{@{}l@{}}Packet corruption \\ (Constant)\end{tabular}}
&30 mins& 60 mins & 172 mins & 148 mins
\\
\hline 
{\it \begin{tabular}[c]{@{}l@{}}Packet corruption \\ (Exponential)\end{tabular}}
&30 mins& 55 mins & 48 mins & 23 mins
\\
\hline 
{\it \begin{tabular}[c]{@{}l@{}}Packet latency (Random)\end{tabular}}
&45 mins& 70 mins & 132 mins & 107 mins
\\
\hline 
{\it \begin{tabular}[c]{@{}l@{}}Packet latency (Constant)\end{tabular}}
&30 mins& 60 mins & 132 mins & 102 mins
\\
\hline 
{\it \begin{tabular}[c]{@{}l@{}}Packet latency \\ (Exponential)\end{tabular}}
&45 mins& 60 mins & 59 mins & 44 mins
\\
\hline 
{\it \begin{tabular}[c]{@{}l@{}}Packet loss (Random)\end{tabular}}
&50 mins& 65 mins & 142 mins & 127 mins
\\
\hline 
{\it \begin{tabular}[c]{@{}l@{}}Packet loss (Constant)\end{tabular}}
&30 mins& 65 mins & 85 mins & 50 mins
\\
\hline 
{\it \begin{tabular}[c]{@{}l@{}}Packet loss \\ (Exponential)\end{tabular}}
&50 mins& 65 mins & 52 mins & 37 mins
\\
\hline 

\end{tabular}
\medskip
\begin{footnotesize}
\emph{$>$12 hours} indicates the cases of failures that have not been observed within 12 hours, although in the presence of active faults that would eventually lead to the system failures.
\end{footnotesize}
\end{center}
\vspace{-0.5cm}
\end{table}

\subsubsection*{RQ5: Comparative Evaluation} 
\label{subsec:rq4}
\begin{change}We compare \approach to both \emph{OA-PI} and \emph{G-BDA} on the same testbed. 
\emph{OA-PI} is a widely adopted industrial anomaly-based tool, \emph{G-BDA} is a state-of-the-art signature-based approach. 
We use \emph{OA-PI} as a baseline approach, and \emph{G-BDA} as a relevant representative of competing approaches.
%
Table~\ref{tab:comparison} reports precision, recall, F-Measure, accuracy and false positive rate of \approach, OA-PI and G-BDA.  \end{change}

OA-PI infers the threshold of normal
performance for KPI values, and raises alarms only for persistent anomalies, that is, if the probability that a KPI is anomalous for 3 of the last 6 intervals is above a certain threshold value~\cite{IBM:ITOAPI:Tutorial:2015}. 
Columns \emph{OA-PI (anomalies)} and \emph{OA-PI (alarms)} of Table~\ref{tab:comparison} report all and the persistent anomalies that OA-PI detects and signal, respectively.  
%
%
In both cases OA-PI is less effective than \approachNoSpace: 
OA-PI does not raise any alarm, thus failing to predict the failure (recall = $0\%$, precision and F-measure not computable), and records far too many anomalies, thus  signalling all potential failures (recall of $100\%$) diluted in myriads false alarms (false positive rate = 100\%). 
In a nutshell, OA-PI reports every legal executions as a possible failure. 
%
\approach is effective: The high values of the five measures indicate that \approach predicts most failures with a negligible amount of false positives.  

\begin{change}
G-BDA is a signature-based approach that collects VM metrics to detect preliminary symptoms of failures.   G-BDA detects both excessive workload and anomalous virtual machines.  G-BDA analyzes a single tier of a distributed system. 
Columns \emph{G-BDA (single-tier)} and \emph{G-BDA (multi-tier)} of Table~\ref{tab:comparison} report precision, recall, F-measure, accuracy and FPR of G-BDA, by referring to the analysis of faults injected in  a single VM and faults injected in different tiers, respectively. In both cases \approach outperforms G-BDA on all five measures, and reduces FPR from over 2\% to 0.05\%.
\end{change}

\begin{change} In summary, the \approach combination of anomaly detection and signature-based analysis is more effective than either of the two techniques used in isolation.
\end{change}

\begin{table}[]
\caption {Comparative evaluation of \approach and state-of-art approches}\label{tab:comparison}
\resizebox{\textwidth}{!}{\begin{tabular}{cccccc}
\hline
\multirow{2}{*}{\textbf{Measures}} & \multirow{2}{*}{\textbf{\approach}} & \multirow{2}{*}{\textbf{\begin{tabular}[c]{@{}c@{}}OA-PI\\ (alarms)\end{tabular}}} & \multirow{2}{*}{\textbf{\begin{tabular}[c]{@{}c@{}}OA-PI \\ (anomalies)\end{tabular}}} & \multirow{2}{*}{\textbf{\begin{tabular}[c]{@{}c@{}}\begin{change}G-BDA\end{change}\\ \begin{change}(single-tier)\end{change}\end{tabular}}} & \multirow{2}{*}{\textbf{\begin{tabular}[c]{@{}c@{}}\begin{change}G-BDA\end{change}\\ \begin{change}(multi-tier)\end{change}\end{tabular}}} \\
                                   &                                    &                                                                                    &                                                                                        &                                                                                         &                                                                                        \\ \hline
Precision                          & 98.798\%                           & --                                                                                 & 94.118\%                                                                               & \begin{change}90.967\%\end{change}                                                                                & \begin{change}87.933\%\end{change}                                                                               \\ \hline
Recall                             & 98.797\%                           & 0\%                                                                                & 100\%                                                                                  & \begin{change}90.567\%\end{change}                                                                                & \begin{change}87.533\%\end{change}                                                                               \\ \hline
F-Measure                          & 98.797\%                           & --                                                                                 & 96.970\%                                                                               & \begin{change}90.367\%\end{change}                                                                                & \begin{change}87.400\%\end{change}                                                                               \\ \hline
Accuracy                           & 98.797\%                           & 5.882\%                                                                            & 94.118\%                                                                               & \begin{change}90.567\%\end{change}                                                                                & \begin{change}87.533\% \end{change}                                                                              \\ \hline
FPR                                & 0.05\%                            & 0\%                                                                                & 100\%                                                                                  & \begin{change}2.833\%\end{change}                                                                                 & \begin{change}2.3\%\end{change}                                                                                  \\ \hline
\end{tabular}}
\end{table}


\subsubsection*{RQ6: Overhead}
\label{subsec:RQ5}

\approach interacts directly with the system only with the \emph{KPI Monitor}, which in our prototype implementation collects the KPIs by means of the SNMPv2c monitoring service for Clearwater~\cite{Case:SNMP:RFC:1996}, the Ceilometer telemetry service for OpenStack~\cite{Openstack:Ceilometer:2015} and a Linux OS agent that we implemented for Ubuntu. All other computation is performed on a dedicated node, and does not impact on the overall performance of the target system.
Thus, the \approach overhead on the running system is limited to the overhead of the monitoring services that we expect to be very low.

\begin{figure}[ht!]
\begin{center}
  \includegraphics[width=10cm]{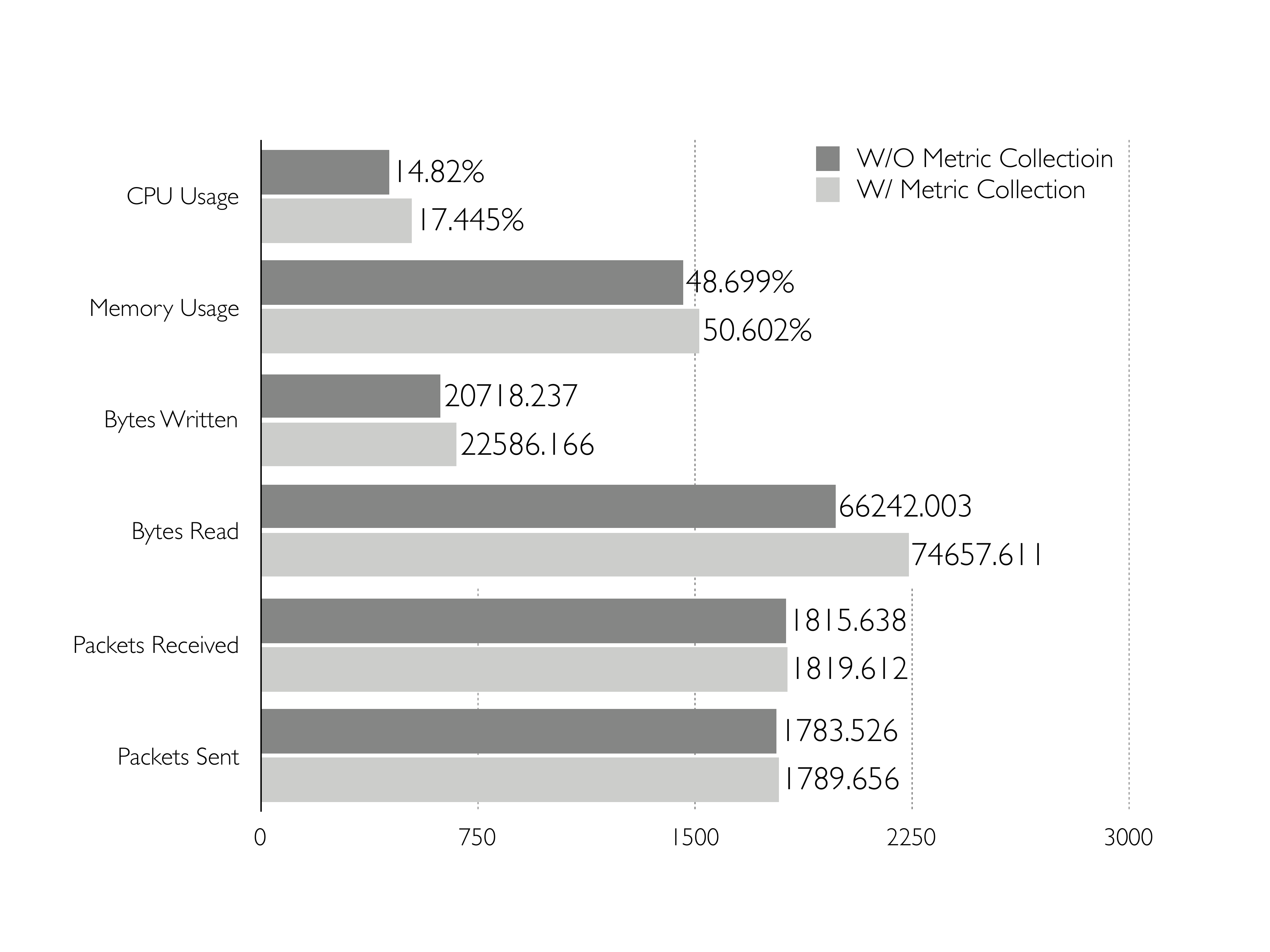}
\caption{\approach overhead}
\label{fig:overhead}
\end{center}
\end{figure}

The experimental results confirm the absence of overhead, which means no measurable difference, on the target system. We only observe small differences in resource consumption as reported in Figure~\ref{fig:overhead}, which reports cpu, memory, disk and network consumption when the system is executed with and without the monitoring infrastructure.
The monitoring infrastructure has a negligible impact on disk (measured as read and written bytes) and network (measured as number of sent and received packets) usage, accounting for few hundreds bytes over tens of thousands and few packets over thousands, respectively.  
The impact on CPU and memory usage is also quite low, with an average increase of 2.63\% and 1.91\%, respectively.
These results are perfectly compatible with systems with strong performance requirements, such as telecommunication infrastructures.

\subsubsection*{Threats To Validity}
\label{sec:threat}

In this article, we reported our experience with an IMS in a network function
virtualization environment.
Although we have achieved success predictors under different runtime conditions,
this may not generalize to other cloud systems.
While the approach is generalised easily, it cannot be assumed in advance that
the results of a study will generalise beyond the specific environment in which
it was conducted. This is an external threat to the validity. \begin{change}One way to
mitigate this threat is to analyze other real-world cloud systems. However,
there is no publicly available benchmark from realistic cloud
deployments such as Project Clearwater. To overcome this limitation, we are partnering with industrial companies to test PreMiSE in their pre-production systems.\end{change}

An internal threat to validity is the limited number of faults we examined in
the study. We chose popular faults without apriori knowledge of their behavior.
However, it is possible that there are faults that do not exhibit any
performance deviations. To mitigate the threat mentioned above, we plan to
extend the study with a larger set of experiments, so that statistical
significant test can be meaningfully applicable.

\section{Related Work}
\label{sec:related}


State-of-the-art techniques for predicting failures and locating the corresponding faults are designed to support system administrators, enable self-healing solutions or dealing with performance issues~\cite{Ibidunmoye:AnomalyDetectionSurvey:2015, Salfner:PredictSurv:ACMCompSurv:2010}. 
Current techniques to predict system failures derive abstractions of the behavior of a system in the form of models and rules, and exploit either \emph{signature-} or \emph{anomaly-based} strategies to dynamically reveal symptoms of problems, and predict related failures.


\emph{Signature-based approaches} capture failure-prone behaviors that indicate how the monitored system behaves when affected by specific faults, and aim to reveal faults of the same kinds at runtime.
\emph{Anomaly-based approaches} capture non-failure-prone behaviors that represent the correct behavior of the monitored system
and aim to reveal behaviors that violate these abstractions at runtime.
\emph{Performance-related approaches} dynamically identify anomalies and bottlenecks. 

Signature-based approaches can accurately predict the occurrences of failures whose symptoms are encoded in the models that capture the fault signatures, but are ineffective against failures that do not correspond to signatures encoded in the models.
Anomaly-based approaches can potentially predict any failure, since they reveal violations of models of the positive behaviors of the application, 
but depend on the completeness of the models, and suffer from many false alarms. 
In a nutshell, \emph{signature-based approaches may miss several failures while anomaly-based approaches  may generate many false alarms}.

\approach integrates anomaly- and signature-based approaches to benefit from both the generality of anomaly-based techniques and the accuracy of signature-based techniques. 
While current signature-based approaches derive failure signatures from application events~\cite{Vilalta:EventPrediction:ICDM:2002,Fu:EventCorrelation:SC:2007,Salfner:PredictingFailure:IPDPS:2006},  the original \approach approach derives failure signatures from anomalies that are good representative of failure-prone behaviors, and thus particularly effective in predicting failures. 

\emph{Performance-related approaches} address performance problems by detecting anomalies and bottlenecks that do not affect the functional behavior of the system, and as such are largely complementary to \approach and related approaches.



\subsection{Signature-Based Approaches}

The main signature-based approaches are
the Vilalta et al.'s approach~\cite{Vilalta:EventPrediction:ICDM:2002}, 
hPrefects~\cite{Fu:EventCorrelation:SC:2007}, 
SEIP~\cite{Salfner:PredictingFailure:IPDPS:2006}, 
Seer~\cite{Ozcelik:Seer:TSE:2016}, 
Sauvanaud et al.~\cite{Sauvanaud:Anomaly:ISSRE:2016},
SunCat~\cite{Nistor:SunCat:ISSTA:2014} 
and the approach defined by Malik et al.~\cite{Malik:AutomaticDetection:ICSE:2013}.

Vilalta et al.\ introduced an approach that mines failure reports to learn associative rules that relate events that frequently occur prior to system failures to the failures themselves, and use the mined rules to predict failures at runtime, before their occurrence~\cite{Vilalta:EventPrediction:ICDM:2002}. 
Fu and Zu's \emph{hPrefects} approach extends Vilalta et al.'s rules to clustered architectures~\cite{Fu:EventCorrelation:SC:2007}. 
\emph{hPrefects} learns how failures propagate in time and space from failure records, represents temporal correlations with a spherical covariance model, and spatial correlations with stochastic models, and includes a cluster-wide failure predictor that uses the learned models to estimate the probability that a failure occurs in the current execution. 

Salfner et al.'s \emph{SEIP} approach synthesizes a semi-Markov chain model that includes information about  error frequency and error patterns~\cite{Salfner:PredictingFailure:IPDPS:2006}, and  signals a possible system failure, when the model indicates that the probability that the current execution will produce a failure exceeds a given threshold. 

Ozchelik and Yilmaz's \emph{Seer} technique combines hardware and software monitoring to reduce the runtime overhead, which is particularly important in telecommunication systems~\cite{Ozcelik:Seer:TSE:2016}. 
\emph{Seers} trains a set of classifiers by labeling the monitored data, such as caller-callee information and number of machine instructions executed in a function call, as passing or failing executions, and uses the classifiers to identify the signatures of incoming failures.

Sauvanaud et al. capture symptoms of service level agreement violations: They collect application-agnostic data, and classify system behaviors as normal and anomalous with a Random Forest algorithm, and show that collecting data from an architectural tier impacts on the accuracy of the predictions~\cite{Sauvanaud:Anomaly:ISSRE:2016}.

Nistor and Ravindranath's \emph{SunCat} approach predicts performance problems in smartphone applications by identifying calling patterns of string getters that may cause performance problems for large inputs, by analyzing similar calling patterns for small inputs~\cite{Nistor:SunCat:ISSTA:2014}.

Malik et al.~\cite{Malik:AutomaticDetection:ICSE:2013} developed an automated approach to detect performance deviations before they become critical problems. The approach collects performance counter variables, extracts performance signatures, and then uses signatures to predict deviations. 
Malik et al. built signatures with a supervised and three unsupervised approaches, and provide experimental evidence that the supervised approach is more accurate than the unsupervised ones even with small and manageable subsets of performance counters.

Lin et al.'s~\cite{Lin:PNF:FSE:2018} \emph{MING} technique uses an ensemble of supervised machine learning models to predict failures in cloud systems by analyzing both temporal and spatial data. Compared to PreMiSe, MING does not consider the multi-level nature of cloud systems and can predict failures only at the granularity of the host. 

 El-Sayed et al.~\cite{ElSayed:LearningFromFailure:ICDCS:2017} note that unsuccessful jobs across different clusters exhibit patterns that distinguish them from successful executions. On the basis of this observation, they use random forests to identify signatures of unsuccessful terminations of jobs or tasks running in the cluster. Predictions at the job level are then used to mitigate the effect of unsuccessful job executions.

\approach introduces several novel features that improve over current signature-based approaches: 
\begin{inparaenum}[(i)]
\item it creates signatures from anomalies, which better represent failure occurrences than general events, 
\item predicts the type of failure that will occur,
\item integrates and correlates data extracted from all layers and components of a multi-tier distributed architecture, and 
\item restricts the scope of the location of the causing faults.
\end{inparaenum}

\subsection{Anomaly-Based Approaches}
The main anomaly-based approaches are the algorithms proposed by Fulp et
al.~\cite{Fulp:PredictingFailure:WASL:2008}, Jin et al.~\cite{Jin:Performance:2007}  and Guan et al.~\cite{Guan:failurePrediction:ICCCN:2011}, and the
\emph{Tiresias}~\cite{Williams:BlackBoxPrediction:IPDPS:2007},
\emph{ALERT}~\cite{Tan:AnomalyPrediction:PODC:2010}, \emph{PREPARE}~\cite{Tan:anomalyPrediction:ICDCS:2012} and \emph{OA-PI}~\cite{IBM:SmartCloudAnalyticsPI:2015} technologies.

Fulp et al.'s approach and \emph{PREPARE} address specific classes of failures.  
Fulp et al. defined a spectrum-kernel support vector machine approach to predict disk failures using system log files~\cite{Fulp:PredictingFailure:WASL:2008},
while \emph{PREPARE} addresses performance anomalies in virtualized systems~\cite{Tan:anomalyPrediction:ICDCS:2012}. 
Fulp et al. exploit the sequential nature of system messages, the message types and the message tags, to distill features, and use support vector machine models to identify message sequences that deviate from the identified features as symptoms of incoming failures.
\emph{PREPARE} combines a 2-dependent Markov model to predict attribute values with a tree-augmented Bayesian network to predict anomalies. 
Differently from both thees approaches that are specific to some classes of failures, \approach is general and can predict multiple types of failures simultaneously. 


Guan et al. proposed an ensemble of Bayesian Network models to characterize the normal execution states of a cloud system~\cite{Guan:failurePrediction:ICCCN:2011}, and to signal incoming failures when detecting states not encoded in the models.
%
\emph{ALERT} introduces the notion of alert states, and exploits a triple-state
multi-variant stream classification scheme to capture special alert states and
generate warnings about incoming failures~\cite{Tan:AnomalyPrediction:PODC:2010}.
%
\emph{Tiresias} integrates anomaly detection and Dispersion Frame Technique (DFT) to predict anomalies~\cite{Williams:BlackBoxPrediction:IPDPS:2007}.
 
Jin et al. use benchmarking and production system monitoring to build an analytic model of the system that can then be used to predict the performance of a legacy system under different conditions to avoid unsatisfactory service levels due to load increases~\cite{Jin:Performance:2007}.



Anomaly-based approaches are inevitably affected by the risk of generating many false positives as soon as novel legal behaviors emerge in the monitored system. \approach overcomes the issue of false positives by integrating an anomaly detection approach with a signature-based technique that issues alarms only when the failure evidence is precise enough.
The results reported in Section~\ref{subsec:rq4} show that \approach dramatically improves over current anomaly-based detection techniques, including modern industrial-level solutions such as IBM \emph{OA-PI}~\cite{IBM:SmartCloudAnalyticsPI:2015}. 

\subsection{Performance-Related Approaches}


Performance anomaly detection approaches predict performance issues and identify bottlenecks in production systems, while performance regression approaches detect performance changes~\cite{Ibidunmoye:AnomalyDetectionSurvey:2015}.

\begin{change}
Classic performance anomaly detection approaches work with historical data: they build statistical models of low-level system metrics to detect performance issues of distributed applications~\cite{Cohen:PerformanceDetection:2005,Bodik:FDA:2010,Lim:PerformanceIssues:14,He:Log:2018}. They derive formal representations, called signatures, that are easy to compute and retrieve, to capture the state of the system, and quickly identify similar performance issues that occurred in the past.  These approaches aim to discriminate different types of performance issues in order to aid root cause analysis. 

The most recent performance anomaly detectors, BARCA, Root and TaskInsight, do not need historical data.\end{change}
BARCA monitors system metrics, computes performance indicators, like mean, standard deviation, skewness and kurtosis, and combines SVMs, to detects anomalies, with multi-class classifier analysis, to identify related anomalous behaviors, like deadlock, livelock, unwanted synchronization, and memory leaks~\cite{Alvarez:BARCA:DSC:2018}. 
Root works as a Platform-as-a-Service (PaaS) extension~\cite{Jayathilaka:ROOT:CC:2018}. 
It detect performance anomalies in the application tier, classifies their cause as either workload change or internal bottleneck, and locates the most likely causes of internal bottlenecks with weighted algorithms. 
TaskInsight detects performance anomalies in cloud applications, by analysing system level metrics, such as CPU and memory utilization, with a clustering algorithm, to identify abnormal application threads\cite{Zhang:TaskInsight:CLOUD:2016}. 
It detects and identifies abnormal application threads by analyzing system level metrics such as CPU and memory utilization. 


 


Differently from \approachNoSpace, these approaches 
\begin{inparaenum}[(i)]
\item do not locate the faulty resource that causes the performance anomaly, and 
\item cannot detect performance problems at different tiers, which remains largely an open challenge~\cite{Ibidunmoye:AnomalyDetectionSurvey:2015}.
\end{inparaenum}

\medskip

\emph{Performance regression} approaches detect changes in software system performance during development aiming to prevent performance degradation in the production system~\cite{Chen:PerformanceRegression:ICSME:2017}. 
They reveal changes in the  overall performance of the development system due to changes in the code. 
Ghaith et al.~\cite{Ghaith:PerformanceRegression:CSMR:2013} detect performance regression by
comparing transaction profiles to reveal performance anomalies that can occur only if the application changes.
Transaction profiles reflects the lower bound of the response time in a transaction under idle condition, and do not depend on the workload. 
Foo et al.~\cite{Foo:PerformanceRegressions:ICSE:2015} detect performance regressions in heterogeneous environments in the context of data centers, by building an ensemble of models to detect performance deviations. 
Foo et al. aggregate performance deviations from different models, by using simple voting as well as weighted algorithms to determine whether the current behavior really deviate from the expected one, and is not a simple  environment-specific variation. 
Performance regression approaches assume a variable system code base and a stable runtime environment, while \approach collects operational data to predict failures and localize faults caused by a variable production environment in an otherwise stable system code base.

\section{Conclusions}
\label{sec:conclusion}

In this paper, we presented \approachNoSpace, an original approach to automatically predict failures and locate the corresponding faults in multi-tier distributed systems, where faults are becoming the norm rather then the exception.
%
%
Predicting failure occurrence as well as locating the responsible faults produce information that is essential for mitigating  the impact of failures and improving the dependability of the systems. 
Current failure prediction approaches rarely produce enough information to locate the faults corresponding to the predicted failures, and either suffer from false positives (anomaly-based) or work with patterns of discrete events and do not cope well with failures that impact on continuous indicators (signature-based). 

\approach originally blends anomaly- and signature-based techniques to address failures that impact on continuous indicators, and to precisely locate the corresponding faults.
It uses data time series analysis and Granger causality tests to accurately reveal anomalies in the behavior of the system as a whole, probabilistic classifiers to distill signatures that can distinguish \begin{review}failure-free\end{review} albeit anomalous behaviors from failure-prone executions, and signature-based techniques to accurately distinguish malign from benign anomalies, predict the type of the incoming failures, and locate the sources of the incoming failures.
\approach executes on a node independent from the target system, and  limits the online interactions with the monitored applications to metric collection.


In this paper, we report the results of experiments executed on the implementation of a Clearwater IP Multimedia Subsystem, which is a system commonly adopted by telecommunication companies for their VOIP (voice over IP), video and message services.
The results confirm that \approach can predict failures and locate faults with higher precision and less false positives than state of the approaches, without incurring in extra execution costs on the target system.
Differently from state-of-the-art approaches, \approach can effectively identify the type of the possible failure and locate the related faults for the kinds of faults and failures used in the training phase.
%


We designed and studied \approach in the context of multi-tier distributed systems, to predict failures and locate faults at the level of individual tier of the nodes of the system. 
Studying the \approach approach in the context of other systems that can be extensively monitored is a promising research direction.

\section{Acknowledgements}
This work has been partially supported by the H2020 Learn project, which has been funded under the ERC Consolidator Grant 2014 program (ERC Grant Agreement n. 646867) and by the GAUSS national research project, which has been funded by the MIUR under the PRIN 2015 program (Contract 2015KWREMX).

\bibliography{main}

\appendix

\newpage
\section{KPI List}
\label{ap:kpi}

Metrics used in the experiments grouped by tier.

\vspace{0.2cm}
\begin{center}
\textit{\textbf{Linux server}}
\end{center}


\begin{footnotesize} \noindent
\begin{tabular}{p{1.4cm} | p{7cm}}
\textbf{CPU} & user cpu utility, system cpu utility, busy cpu utility, wait io cpu utility\\
\hline
\textbf{Network} & bytes received per sec, bytes transmitted per sec\\
\hline
\textbf{System} & context switches per sec, pages swapped in per sec, pages swapped out per sec, pages faults per sec, total num of processes per sec\\
\hline
\textbf{Virtual Memory} & percentage of swapped space used, percentage of memory used, percentage of memory in buffers, percentage of memory cached.\\
\hline
\textbf{Socket} & sockets in use\\
\hline
\textbf{I/O} & device name, avg wait time, avg request queue length, read bytes per sec, write bytes per sec\\
\end{tabular}
\end{footnotesize}

\newpage
\begin{center}
\textit{\textbf{OpenStack}}
\end{center}

\begin{footnotesize} \noindent
\begin{tabular}{p{1.4cm} | p{7cm}}
\textbf{Compute} & 
    existence of instance and instance type,
    allocated and used RAM in MB,
    CPU time used,
    avg CPU utilisation,
    num of VCPUs,
    num of read and write requests,
    avg rate of read and write requests per sec,
    volume of reads and writes in Byte,
    avg rate of reads and writes in Byte per sec,
    num of incoming bytes on a VM network interface,
    avg rate per sec of incoming bytes on a VM network interface,
    num of outgoing bytes on a VM network interface,
    avg rate per sec of outgoing bytes on a VM network interface,
    num of incoming packets on a VM network interface,
    avg rate per sec of incoming packets on a VM network interface,
    num of outgoing packets on a VM network interface,
    avg rate per sec of outgoing packets on a VM network interface\\
%
\hline
\textbf{Network} &
    existence of network,
    creation requests for this network,
    update requests for this network,
    existence of subnet,
    creation requests for this subnet,
    update requests for this subnet,
    existence of port,
    creation requests for this port,
    update requests for this port,
    existence of router,
    creation requests for this router,
    update requests for this router,
    existence of floating ip,
    creation requests for this floating ip,
    update requests for this floating ip\\
\hline
\textbf{Controller} &
    image polling,
    uploaded image size,
    num of update on the image,
    num of upload of the image,
    num of delete on the image,
    image is downloaded,
    image is served out\\
\end{tabular}
\end{footnotesize}

\newpage
\begin{center}
\textit{\textbf{\textbf{Clearwater}}}
\end{center}
%
\begin{footnotesize} \noindent
\begin{tabular}{p{2.5cm} | p{5.6cm}}

\textbf{CPU and Memory} & SNMP CPU and memory usage stats\\ 
\hline

\textbf{Latency} & avg latency, variance, highest call latency, lowest call latency\\
\hline

\textbf{Incoming Requests} & num of incoming requests\\
\hline

\textbf{Rejected Requests} & num of requests rejected due to overload \\
\hline

\textbf{Queue} & avg request queue size, variance, highest queue size, lowest queue\\
\hline

\textbf{Cx Interface} & avg latency, variance, highest latency and lowest latency seen on the Cx interface\\
\hline

\textbf{Multimedia-Auth Requests} & avg latency, variance, highest latency and lowest latency seen on Multimedia-Auth Requests\\ 
\hline

\textbf{Server-Assignment Requests} & avg latency, variance, highest latency and lowest latency seen on Server-Assignment Requests\\ 
\hline

\textbf{User-Authorization Requests} & avg latency, variance, highest latency and lowest latency seen on User-Authorization Requests\\ 
\hline

\textbf{Location-Information Requests} & avg latency, variance, highest latency and lowest latency seen on Location-Information Requests\\
\hline

\textbf{Sprout} & avg latency, variance, highest and lowest latency between Sprout and Homer XDMS\\
\hline

\textbf{TCP Connections} & num of parallel TCP connections\\
\end{tabular}
\end{footnotesize}

%


\end{document}